\documentclass[aps,prl,onecolumn,superscriptaddress,showpacs,preprintnumbers,amsmath,amssymb]{revtex4}
\pdfoutput=1

\usepackage{graphicx} % Include figure files
\usepackage{dcolumn}  % Align table columns on decimal point

\graphicspath{{ps}}

\newcommand{\ee}{e^{+}e^{-}}

\newcommand{\jp}{J/\psi}

\newcommand{\pipi}{\pi^{+}\pi^{-}}

\newcommand{\pim}{\pi^{-}}
\newcommand{\pip}{\pi^{+}}

\newcommand{\ppbar}{p\bar{p}}
\newcommand{\uubar}{u\bar{u}}
\newcommand{\ddbar}{d\bar{d}}
\newcommand{\ssbar}{s\bar{s}}
\newcommand{\ccbar}{c\bar{c}}
\newcommand{\bbbar}{b\bar{b}}

\newcommand{\qqbar}{q\bar{q}}
\newcommand{\QQbar}{Q\bar{Q}}

\newcommand{\rt}{\rightarrow}

\newcommand{\etal}{\em et al.}

\newcommand{\lm}{\Lambda}

\newcommand{\lmb}{\bar{\Lambda}}

\newcommand{\yones}{\Upsilon(1S)}
\newcommand{\ytwos}{\Upsilon(2S)}

\begin{document}

%\vspace*{-3\baselineskip}
%\resizebox{!}{3cm}{\includegraphics{belle.eps}}

%\preprint{\vbox{ 
%%                 \hbox{BELLE-CONF-0541}
%                   \hbox{\today} 
%%                 \hbox{hep-ex nnnn, if available}
%}}

\title{ \quad\\[0.5cm] QCD Exotics }

%%%% >>>>> insert the authorlist here. BEFORE the abstract !!!!! <<<<<
%\input author-conf2005

\author{Stephen Lars Olsen}

\affiliation{Seoul National University, Seoul} % Seoul

\date{\today}

%\collaboration{Belle Collaboration}
%\noaffiliation

\begin{abstract}
QCD-motivated models for hadrons predict an assortment of ``exotic" hadrons that have structures that are more
complex then the quark-antiquark mesons and three-quark baryons of the original quark-parton model.  These
include pentaquark baryons, the six-quark $H$-dibaryon, and tetra-quark, hybrid, and glueball mesons.  Despite extensive
experimental searches, no unambiguous candidates for any of these exotic configurations have yet to be identified.
On the other hand, a number of meson states, one that seems to be a proton-antiproton bound state, and others
that contain either charmed-anticharmed quark pairs or bottom-antibottom quark pairs, have been recently discovered
that neither fit into the quark-antiquark meson picture nor match the expected properties of the QCD-inspired exotics.
Here I briefly review results from a recent search for the $H$-dibaryon, and discuss some properties of the
newly discovered states --the so-called $XYZ$ mesons-- and compare them with expectations for conventional
quark-antiquark mesons and the predicted QCD-exotic states.  
\end{abstract}

\pacs{14.40.Gx, 12.39.Mk, 13.20.He}

\maketitle

%%%% >>>> keep the final version single-spaced
%\tightenlines

{\renewcommand{\thefootnote}{\fnsymbol{footnote}}}
\setcounter{footnote}{0}

\section{Introduction}
\noindent
The strongly interacting particles of the Standard Model (SM) are colored quarks
and gluons, the strongly interacting particles in nature are color-singlet
mesons and baryons.  In the SM, quarks and gluons are related to mesons
and baryons by the long-distance regime of QCD, which remains the least understood
aspect of the theory.  Since first-principle lattice-QCD (LQCD) calculations are still
not practical for most long-distance phenomena, a number of models motivated by
the color structure of QCD have been proposed.  However, so far at least, 
predictions of these QCD-motivated models that pertain to the spectrum of hadrons
have not had great success.

\begin{figure}[htb]
%Figure with side by side by side panels
  \includegraphics[height=0.85\textwidth,width=0.7\textwidth]{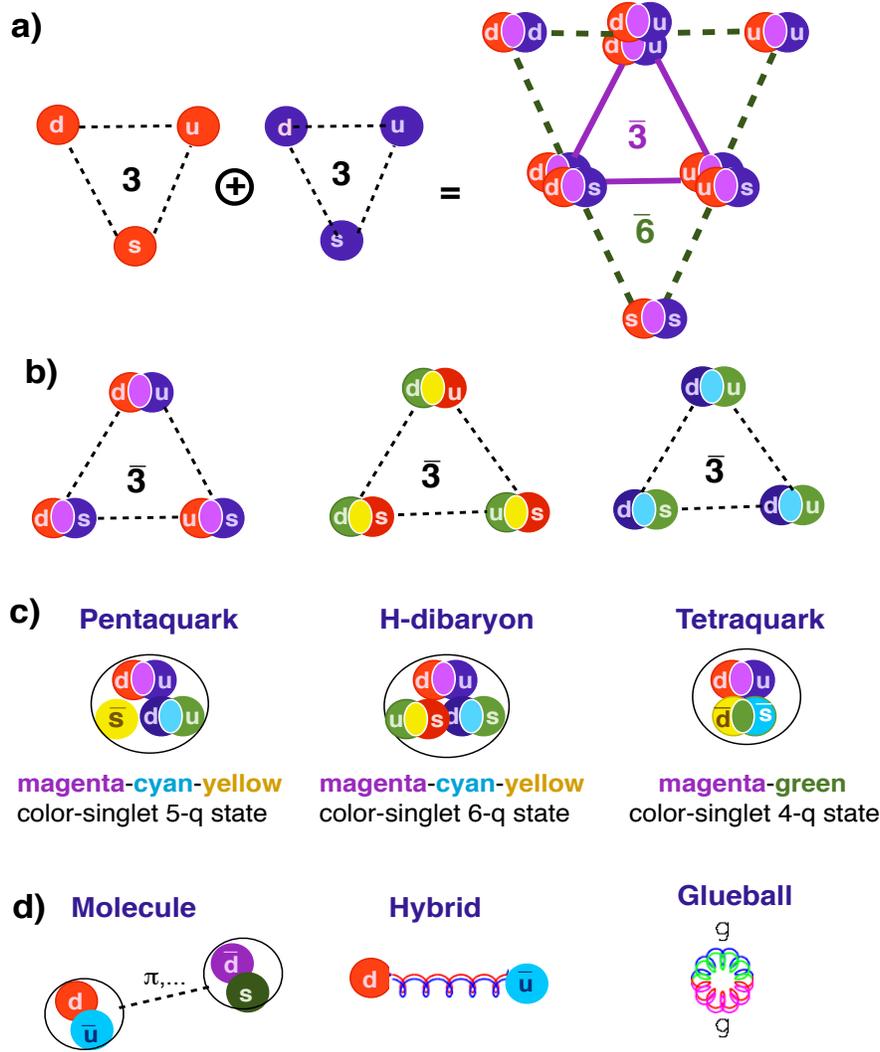}
\caption{\footnotesize {\bf a)} Combining a red and blue quark triplet produces a
magenta (antigreen) antitriplet and a sextet. The antitriplet contains antisymmetric 
$qq$ combinations, the sextet has all symmetric combinations. {\bf b)}  The three anticolored diquark antitriplets.
{\bf c)}  Some of the multiquark, color-singlet states that can be formed from quarks,
antiquarks, diquarks and diantiquarks.
{\bf d)} Other possible non-$\qqbar$ meson systems.}
\label{fig:diquarks}
\end{figure}

For example, it is well known that combining a quark triplet with an antiquark antitriplet
gives the familiar meson octet of flavor-$SU(3)$.  Using similar considerations based
on QCD,  two quark triplets can be combined to form a ``diquark'' antitriplet and
a sextet as illustrated in Fig.~\ref{fig:diquarks}a.  In QCD, these diquarks will have color: 
combining a red triplet with a blue triplet -- as shown in the figure -- produces a magenta
(anti-green) diquark and for the antitriplet combination, which is antisymmetric in color space, 
the color force between the two quarks is expected to be attractive.  Likewise, green-red and blue-green
diquarks form yellow (antiblue) and cyan (antired) antitriplets as shown in Fig.~\ref{fig:diquarks}b.

Since these diquarks are not color-singlets, they cannot exist as free particles but, on the
other hand, the  anticolored diquark antitriplets should be able to combine with other colored
objects in a manner similar to antiquark antitriplets, thereby forming multiquark color-singlet
states with more a complex substructure than the $\qqbar$ mesons and $qqq$ baryons of the original
quark model. These so-called ``exotic'' states  include pentaquark baryons, $H$-dibaryons
and tetraquark mesons, and are illustrated in Fig.~\ref{fig:diquarks}c.
Other proposed exotic states are glueballs, which are mesons made only from gluons,
hybrids formed from a $q$, $\bar{q}$ and a gluon, and deuteron-like bound states of
color-singlet ``normal'' hadrons, commonly referred to as molecules. These
are illustrated in Fig.~\ref{fig:diquarks}d.   Glueball and hybrid mesons are motivated by QCD;
molecules are a generalization of classical nuclear physics to systems of subatomic partics.

\section{\boldmath Pentaquarks and $H$-dibaryons}
All of the above-mentioned candidates for exotic states have been the subject
of numerous theoretical and experimental investigations during the four decades
that have elapsed since QCD was first formulated. This activity peaked in 2003 when
the LEPS experiment at the SPRING-8 electron ring in Japan reported the observation
of a peak in the $K^+ n$ invariant mass distribution in $\gamma n\rt K^+K^-$ reactions
on a carbon target~\cite{leps}, with properties close to those that had been predicted
for the $S=+1$ $\Theta_5^+$ pentaquark~\cite{diakonov}.  This created a lot of excitement
at the time~\cite{schumacher} but subsequent, high-statistics experiments~\cite{penta}
gave negative results.   The current ``common wisdom'' is that pentaquarks do not
exist~\cite{trilling},  or at least have yet to be found.

The $H$-dibaryon was predicted by Jaffee in 1977 to be a doubly strange, tightly
bound six-quark structure $(uuddss)$ with isospin zero and $J^P=0^+$~\cite{jaffe_H}.  
An $S=-2$ state with baryon number $B=2$ and mass below $2m_{\lm}$ could only decay via
weak interactions and, thus, would be long-lived.  Although Jaffe's original prediction
that the $H$ would be $\sim 80$~MeV below the $2m_{\lm}$ threshold was ruled
out by the observation of double-$\lm$ hypernuclei, most notably the famous
``Nagara'' event~\cite{nagara} that limited the allowed $H$ region to masses above
$2m_{\lm}-7.7$~MeV, the theoretical case for an $H$-dibaryon with mass near $2m_{\lm}$
continues to be strong, and has been recently strengthened by two independent LQCD
calculations, both of which find an $H$-dibaryon state with mass near $2m_{\lm}$~\cite{LQCD}.

\subsection{\boldmath Belle $H$-dibaryon search}
The Belle experiment recently reported results of a search for production of
an $H$-dibaryon with mass near $2m_{\lm}$ in inclusive $\yones$ and $\ytwos$
decays~\cite{belle_H}.  Decays of narrow $\Upsilon(nS)$ $(n=1,2,3)$ bottomonium $(\bbbar)$
resonances are particularly well suited for searches for multiquark states 
with non-zero strangeness.  The $\Upsilon(nS)$ states are flavor-$SU(3)$
singlets that primarily decay via annihilation into three gluons.  The gluons
materialize into $\uubar$, $\ddbar$ and $\ssbar$ pairs with nearly equal probabilities,
creating final states with a high density of quarks and antiquarks in a limited volume of
phase space.  A benchmark for the rate for multiquark-state production in these
decays is the measured inclusive decay branching fractions to antideuterons
$\bar{D}$: ${\mathcal B}(\yones\rt \bar{D} + X)=(2.9\pm 0.3) \times 10^{-5}$ and
${\mathcal B}(\ytwos\rt \bar{D} + X)=(3.4\pm 0.6) \times 10^{-5}$~\cite{PDG}.  
If the six-quark $H$-dibaryon is produced at a rate that is similar to that for six-quark antideuterons,
there should be many thousands of them in the 102~million $\yones$ and 158~million
$\ytwos$ event samples collected by Belle.

\begin{figure}[htb]
%Figure with side by side by side panels
\begin{minipage}[t]{44mm}
  \includegraphics[height=0.6\textwidth,width=1.0\textwidth]{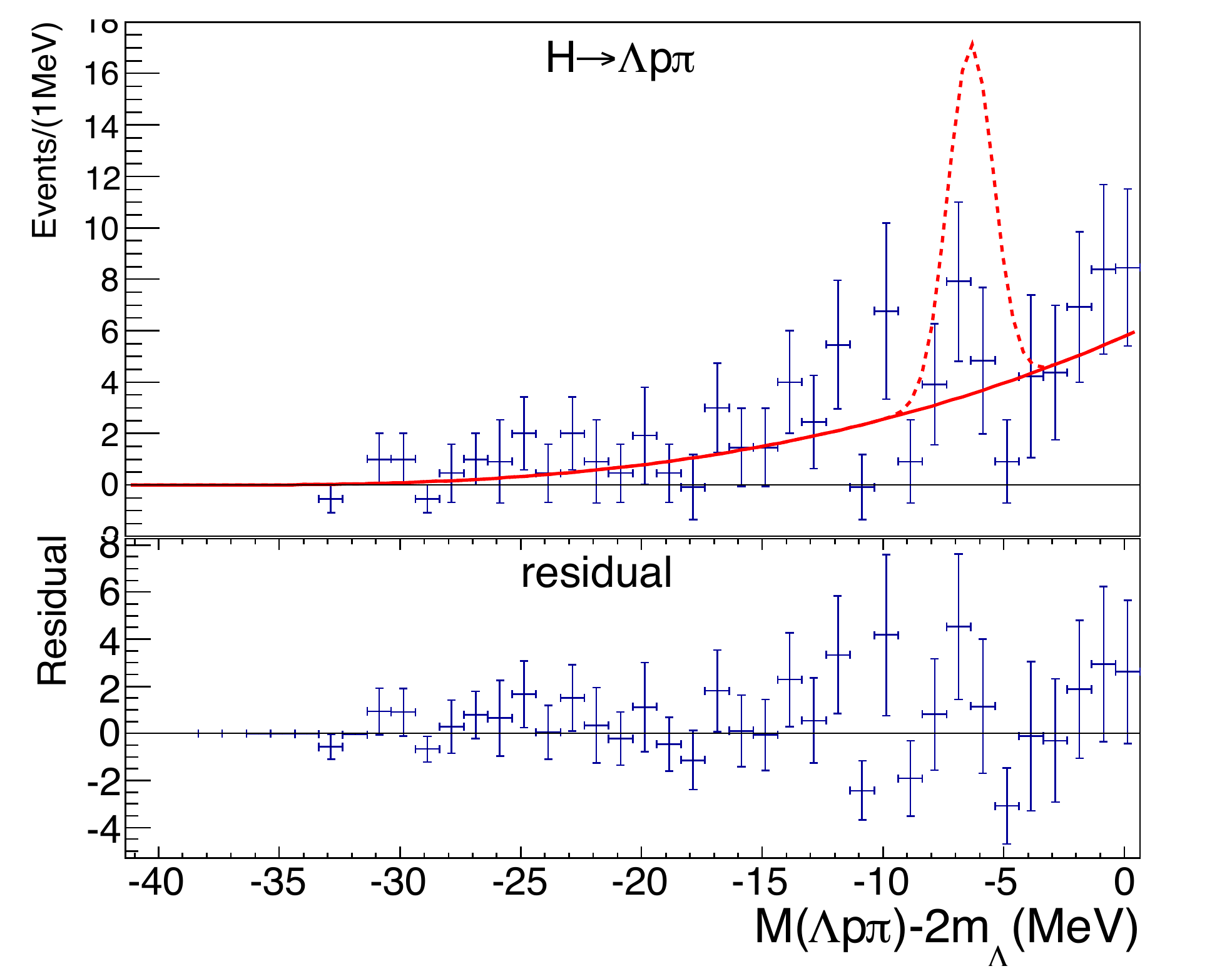}
\end{minipage}
\begin{minipage}[t]{44mm}
  \includegraphics[height=0.6\textwidth,width=1.0\textwidth]{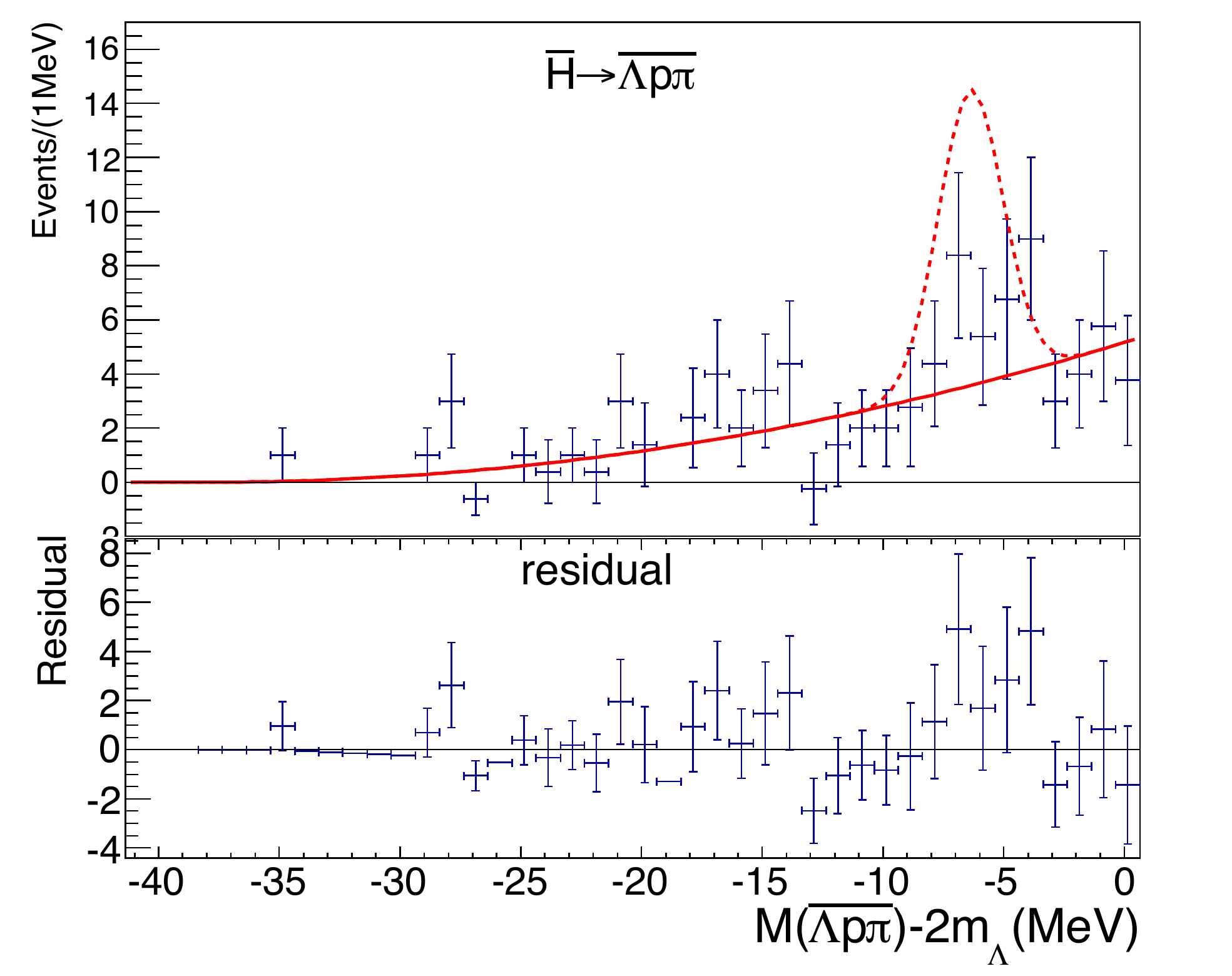}
\end{minipage}
\begin{minipage}[t]{44mm}
  \includegraphics[height=0.6\textwidth,width=1.0\textwidth]{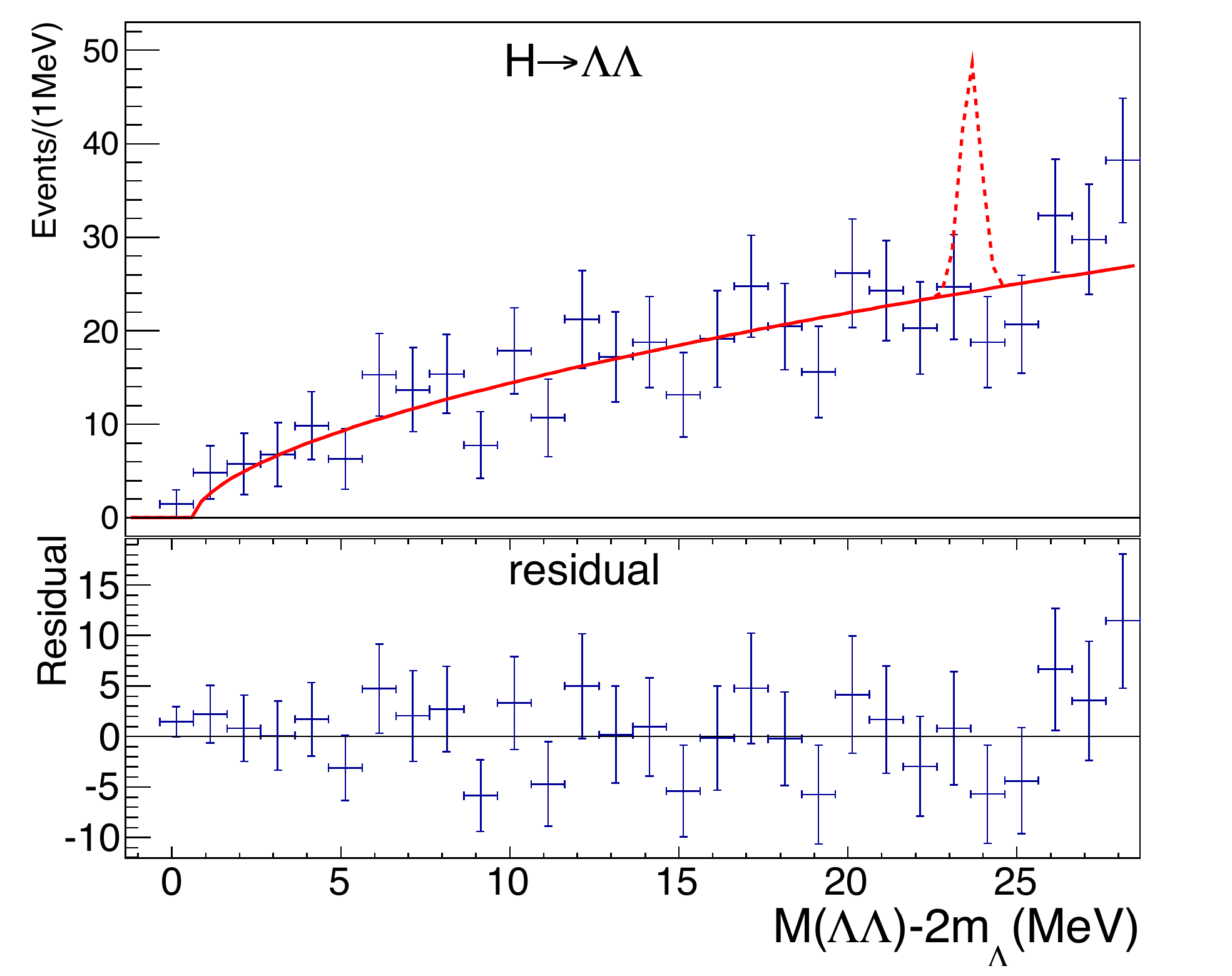}
\end{minipage}
\begin{minipage}[t]{44mm}
  \includegraphics[height=0.6\textwidth,width=1.0\textwidth]{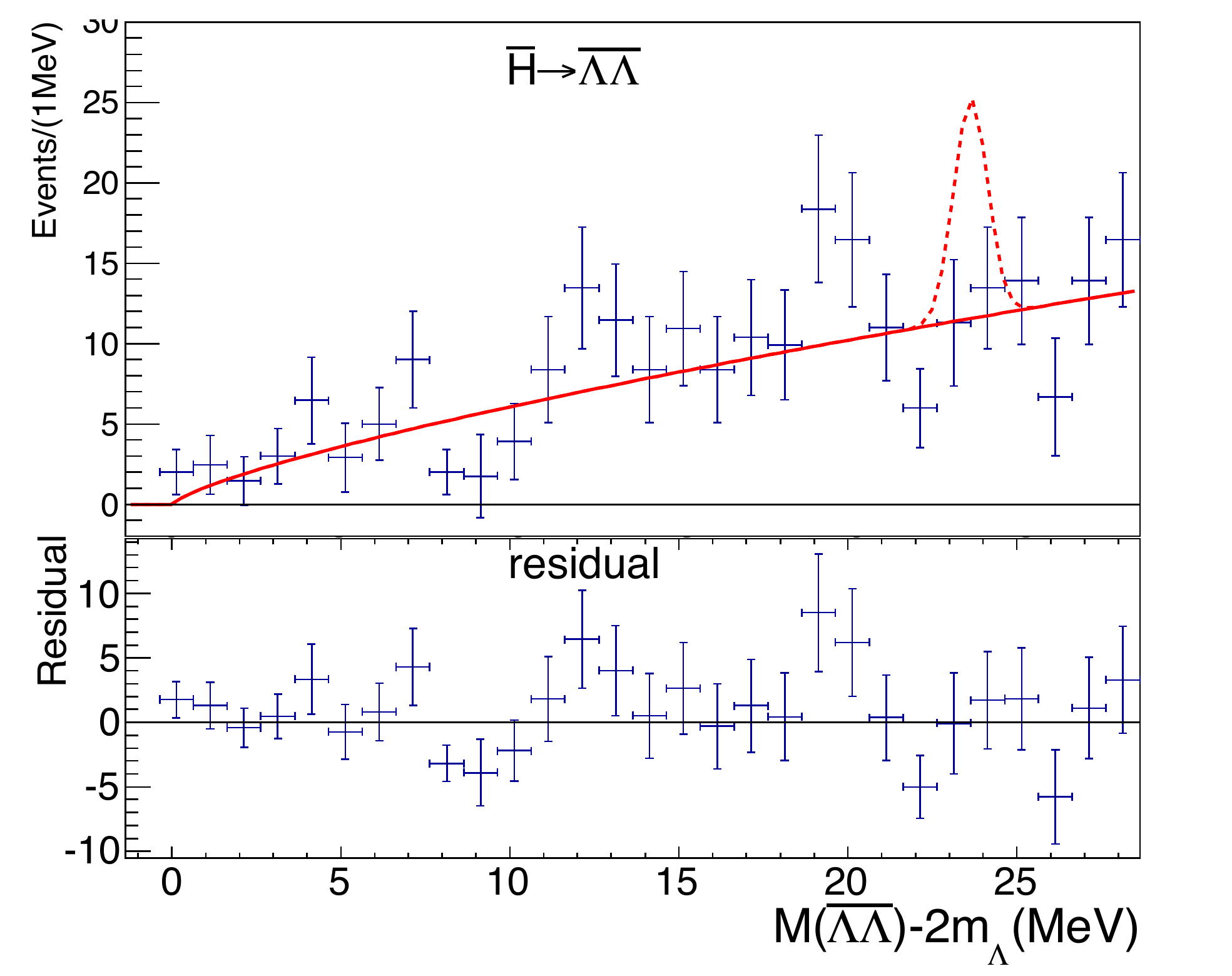}
\end{minipage}\hspace{\fill}
\caption{\footnotesize The $\lm p\pim$, $\bar{\lm}\bar{p}\pip$, $\lm\lm$ and $\bar{\lm}\bar{\lm}$
invariant mass distributions (from left to right).  The solid curves show results of background-only fits
and the lower panels show the fit residuals.  The dashed curves
are expected signals for an $H$ production rate that is 1/20th that for $\bar{D}$'s. 
}
\label{fig:H-plots}
\end{figure}

For $H$ masses below $2m_{\lm}$, Belle searched for $H\rt\lm p\pim$ 
(\& $\bar{H}\rt \bar{\lm}\bar{p}\pip$) signals in the inclusive
$\lm p\pim$ invariant mass distribution~\cite{conj}.  For masses
above $2m_{\lm}$, the $H\rt \lm\lm$ (\& $\bar{H}\rt\bar{\lm}\bar{\lm})$
mode was used.  Figure~\ref{fig:H-plots} shows the measured $\lm p\pim$ (left)
\& $\lmb \bar{p}\pip$ (center-left) invariant mass spectra for masses below $2m_{\lm}$
and the $\lm\lm$ (center-right) and $\lmb\lmb$ (right) mass spectra for
masses above $2m_{\lm}$. Here results from the $\yones$ and $\ytwos$ data samples
are combined.  No signal is observed.  The solid red curves show results of a
background-only fit to the data; the dashed curve shows the MC expectations for
an $H$-dibaryon produced at $1/20$th of the antideuteron rate.
Upper limits on the inclusive branching ratios that are at least a factor of
twenty below that for antideuterons are set over the entire $|M_{H}-2m_{\lm}|<30$~MeV
mass interval. 

Neither pentaquarks nor the $H$-dibaryon are seen in spite of the stong theoretical motivation for their
existence.  The absence of pentaquarks led Wilczek to remark ``{\it The story of the pentaquark shows how poorly
we understand QCD}''~\cite{wilczek_quote}.  
The absence of any evidence for the $H$-dibaryon (among other things) led Jaffe to observe that ``{\it The absence
of exotics is one of the most obvious features of QCD}''~\cite{jaffe_quote}.

\section{What we do see}
\noindent
Although forty years of experimental searches has failed to come up with compelling evidence for
specifically  QCD-motivated exotic hadrons, strong evidence for mesons that do not fit into the simple $\qqbar$ scheme
of the original quark model has been steadily accumulating during the past decade.  These include a candidate for
a bound state of a proton and antiproton from the BESII experiment~\cite{bes2_gppb} -- so-called ``baryonium''-- which
is an idea that has been around for a long, long time~\cite{fermi-yang}, and the $XYZ$ mesons, which are charmonium-like
and bottomonium-like states that do not fit into any of the remaining unfilled states in the $\ccbar$- and $\bbbar$-meson
level schemes~\cite{godfrey-olsen}.

\subsection{\boldmath Baryonium in radiative $\jp\rt\gamma\ppbar$ decays?}

\begin{figure}[htb]
%Figure with side by side by side panels
\begin{minipage}[t]{44mm}
  \includegraphics[height=0.7\textwidth,width=1.2\textwidth]{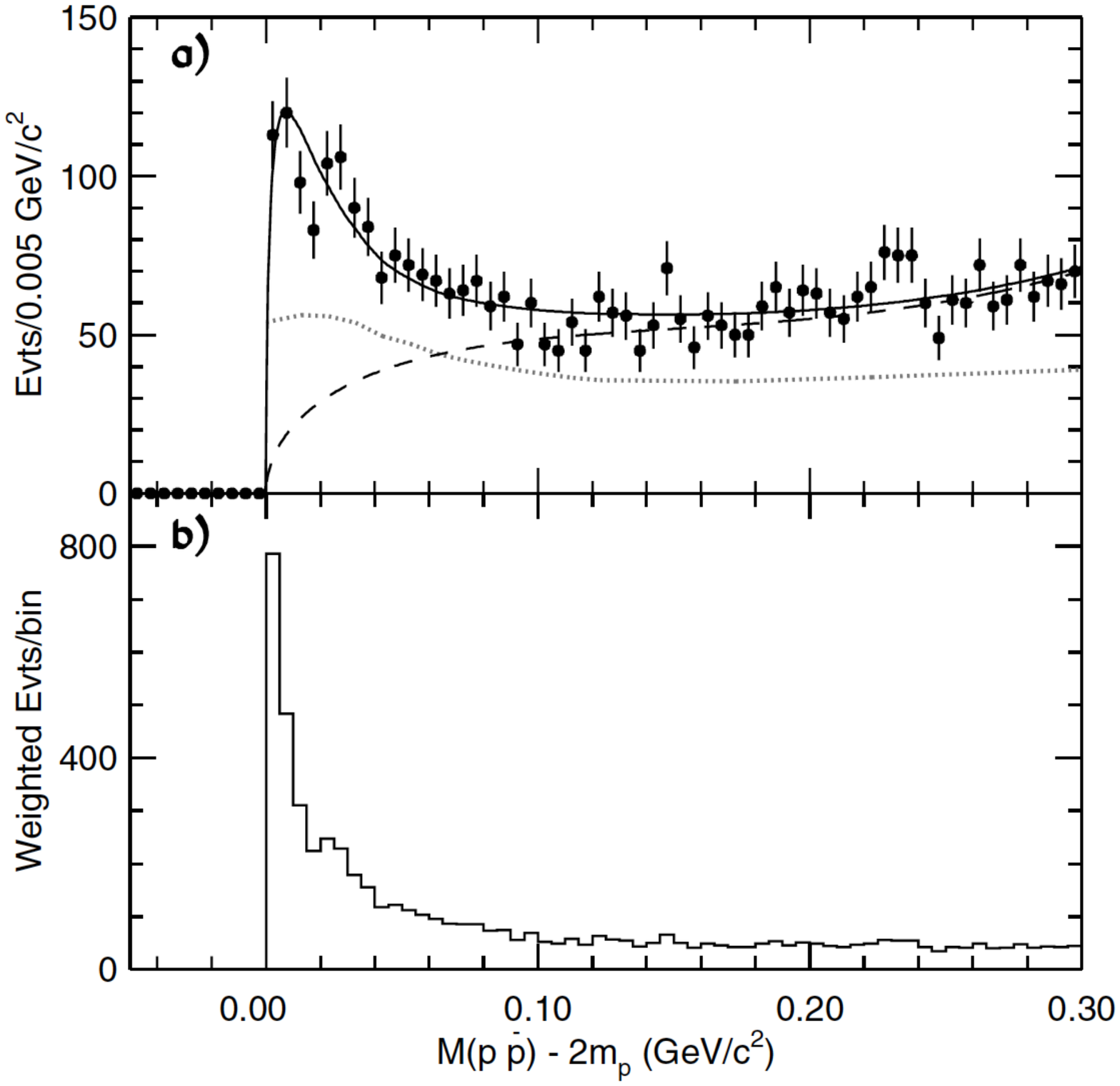}
\end{minipage}
\begin{minipage}[t]{44mm}
  \includegraphics[height=0.7\textwidth,width=1.2\textwidth]{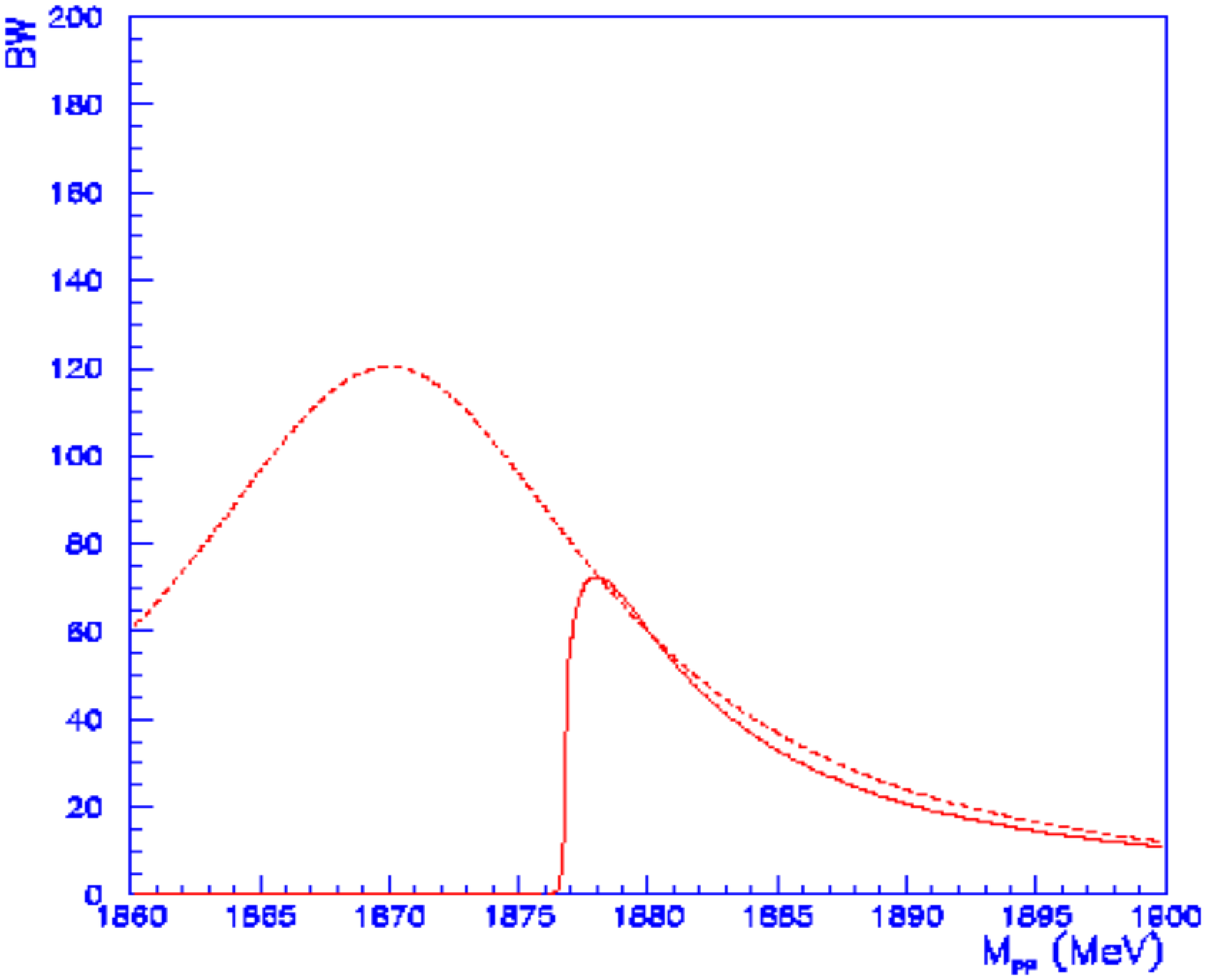}
\end{minipage}
\begin{minipage}[t]{44mm}
  \includegraphics[height=0.7\textwidth,width=1.2\textwidth]{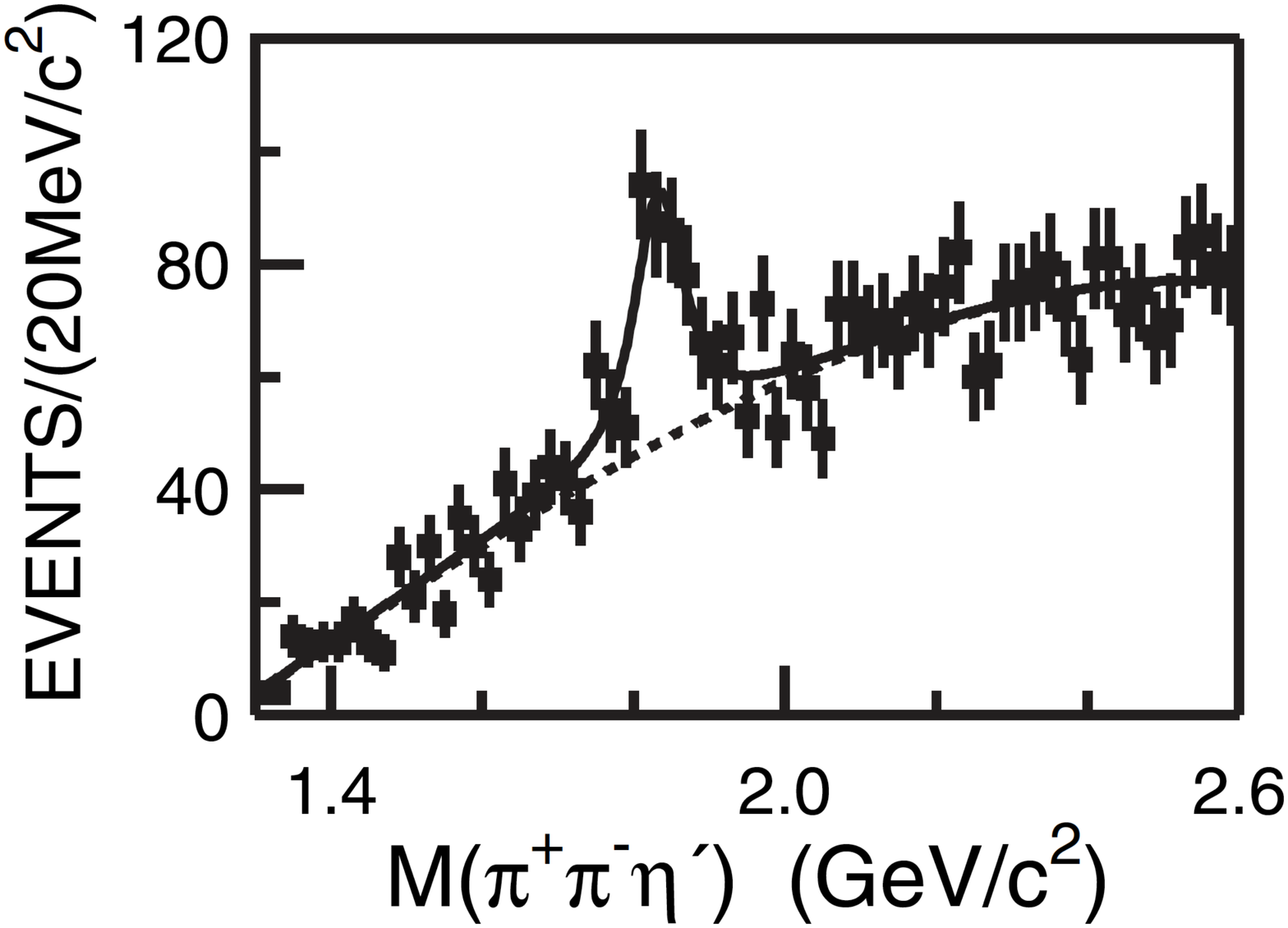}
\end{minipage}
\begin{minipage}[t]{44mm}
  \includegraphics[height=0.7\textwidth,width=1.2\textwidth]{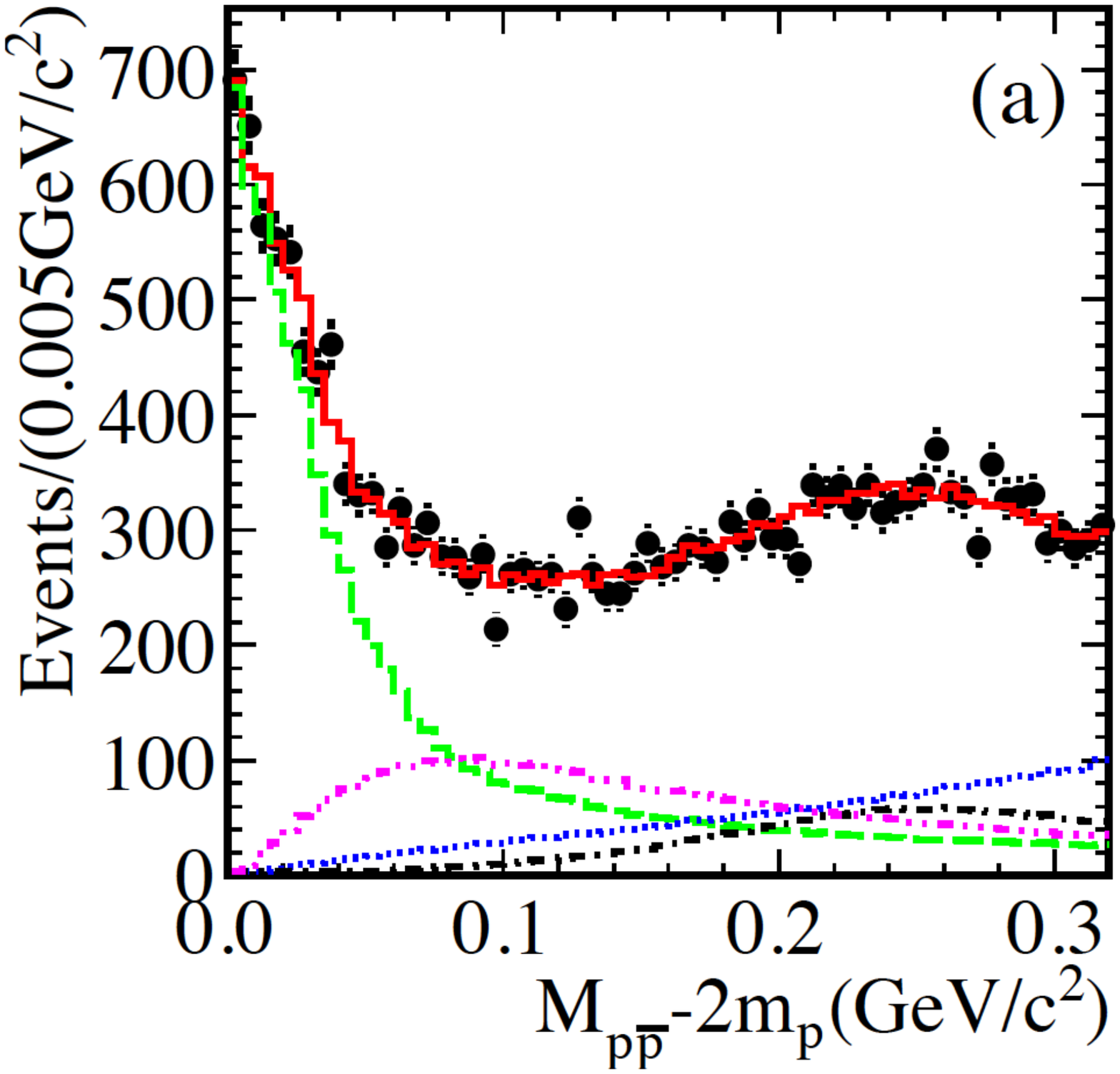}
\end{minipage}
\hspace{\fill}
\caption{\footnotesize {\bf Left)} The upper panel shows the $M(\ppbar)$ distribution
for $\jp\rt\gamma\ppbar$ decays from BESII~\cite{bes2_gppb}.
The lower panel shows the same distribution with the kinematic
threshold suppression factor removed.
{\bf Center-left)} A cartoon showing a BW line shape and a threshold-attenuated BW line-shape
for a hypothesized baryonium state.
The solid curve shows the expected $\ppbar$ line shape.  Below threshold, where the dashed curve
dominates, the state decays via $\ppbar$ annihilation.
{\bf Center-right)} The $M(\pipi\eta^{\prime}$ mass distribution for $\jp\rt\gamma\pipi\eta^{\prime}$
decays from ref.~\cite{bes2_x1835}. 
{\bf Right)} The $M(\ppbar)$ distribution for $\jp\rt\gamma\ppbar$ from BESIII~\cite{bes3_gppb}
with PWA results shown as histograms.}
\label{fig:bes2_gppb}
\end{figure}

In 2003, the BESII experiment reported the observation of a strong near-threshold mass enhancement
in the $\ppbar$ invariant mass spectrum in radiative $\jp\rt\gamma\ppbar$ decays, shown
in the top-left panel of Fig.~\ref{fig:bes2_gppb}~\cite{bes2_gppb}. The
lower panel in the figure shows how the $M(\ppbar)$ spectrum looks when the effects of
phase-space are divided out (assuming an $S$-wave $\ppbar$ system).  It seems apparent
from the phase-space-corrected plot that the dynamical source for this enhancement,
whatever it may be, is at or below the mass threshold.  A fit with a Breit-Wigner (BW) line
shape modified by a kinematic threshold factor yielded a peak mass of $1859^{+6}_{-27}$~MeV
and width $\Gamma<30$~MeV~\cite{errors}.  It was subsequently pointed out that the BW form used by
BESII should be modified to include the effect of final-state-interactions on the shape
of the $\ppbar$ mass spectrum~\cite{fsi}.  When this is done, the peak mass shifts downward,
from 1859~MeV to $1831\pm 7$~MeV and the range of allowed widths increases to
$\Gamma<153$~MeV 

Soon after the BESII publication,  Yan and Ding proposed a Skyrme-like model for
proton-antiproton interactions in which BESII's $\ppbar$ mass-threshold enhancement is an
$S$-wave $p\bar{p}$ bound state with binding energy around 20~MeV~\cite{yan-ding}. Since
the $p$ and the $\bar{p}$ in such a system would annihilate whenever they came within close
proximity of each other, such a state would have a finite width, creating a situation illustrated
by the cartoon in the center-left panel of Fig.~\ref{fig:bes2_gppb}: for masses above the $2m_p$
threshold, the state would decay essentially 100\% of the time by ``falling apart'' into a $p$ and
$\bar{p}$; for masses below $2m_{p}$, the decay would proceed via $p\bar{p}$ annihilation.  Since
a preferred channel for low-energy $S$-wave $\ppbar$ annihlation is $\pipi\eta^{\prime}$,
Yan and Ding advocated a search for subthreshold decays of this same state in radiative
$\jp\rt\gamma\pipi\eta^{\prime}$ decays.  A subsequent BESII study of $\jp\rt\gamma\pipi\eta^{\prime}$
decays found a distinct peak at $1834\pm 7$~MeV and width $68\pm 21$~MeV as shown in
the center-right panel of Fig.~\ref{fig:bes2_gppb}~\cite{bes2_x1835}, in good agreement
with the mass and width results from the FSI-corrected fit to the $p\bar{p}$ mass spectrum.

However, the situation still remains unclear.  The $\ppbar$ mass-threshold enhancement was
confirmed at the same mass with much higher statistics by BESIII~\cite{bes3_gppb}, now with a
significance that is $>30\sigma$ (see the
right panel of Fig.~\ref{fig:bes2_gppb}). The large BESIII event sample permitted the application of
a partial wave analysis (PWA) that established the $J^{PC}=0^{-+}$ quantum number assignment, in
agreement with baryonium expectations.  The $M(\pipi\eta^{\prime})$ peak in $\jp\rt\pipi\eta^{\prime}$
decays was also confirmed and the production angle distribution was found to be consistent with
a $J^{PC}=0^{-+}$ assignment. But the new result finds a much larger width than that
found for the $\ppbar$ peak in the BESIII partial wave analysis: 
$\Gamma_{\pipi\eta^{\prime}}=190\pm 38$~MeV versus the $\Gamma_{\ppbar}<76$~MeV upper limit from the BESIII
PWA. Another puzzling feature is the lack of any evidence for the $\ppbar$ threshold enhancement
in any other channels, such as $\jp\rt\omega\ppbar$~\cite{bes2_wppb},
$\yones\rt\gamma\ppbar$~\cite{cleo_u1s_gppb}  or in $B$ decays~\cite{belle_ppb}.  BESIII is
actively looking at various other radiative $\jp$ decay channels for evidence for or against
a resonance near 1835~MeV~\cite{bes3_g6pi}.

\subsection{\boldmath The $XYZ$ mesons}

The $XYZ$ mesons are a class of hadrons that are seen to decay to final states that contain 
a heavy quark and a heavy antiquark (i.e., $Q$ and $\bar{Q}$, where $Q$ is either a $c$ or $b$
quark), but cannot be easily accommodated in an unfilled $\QQbar$ level.  Since the $c$ and $b$
quarks are heavy, their production from the vacuum in the fragmentation process is heavily suppressed.
Thus, heavy quarks that are seen among the decay products of a hadron must have existed among
its original constituents.  In addition, the heavy quarks in conventional $\QQbar$ ``quarkonium'' mesons
are slow and can be described reasonably well by non-relativistic Quantum Mechanics.  Indeed it
was the success of the charmonium model description of the $\psi$ and $\chi_c$ states in the
mid-1970's that led to the general acceptance of the reality of quarks and the validity of the
quark model.   The Quarkonium model specifies the allowed states of a $\QQbar$ system; if a meson
decays to a final state with a $Q$ and a $\bar{Q}$ and does not match the expected properties
of any of the unfilled levels in the associated quarkonium spectrum, it is necessarily exotic.

\subsubsection{Charmoniumlike mesons}
Charmoniumlike $XYZ$ mesons were first observed in 2003 and continue to be found at a rate of
about one or two new ones every year. There is a huge theoretical and experimental literature on this subject
that cannot be reviewed here.  Instead I restrict myself to a few general remarks.

\begin{figure}[htb]
%Figure with side by side by side panels
\begin{minipage}[t]{85mm}
  \includegraphics[height=1.0\textwidth,width=1.0\textwidth]{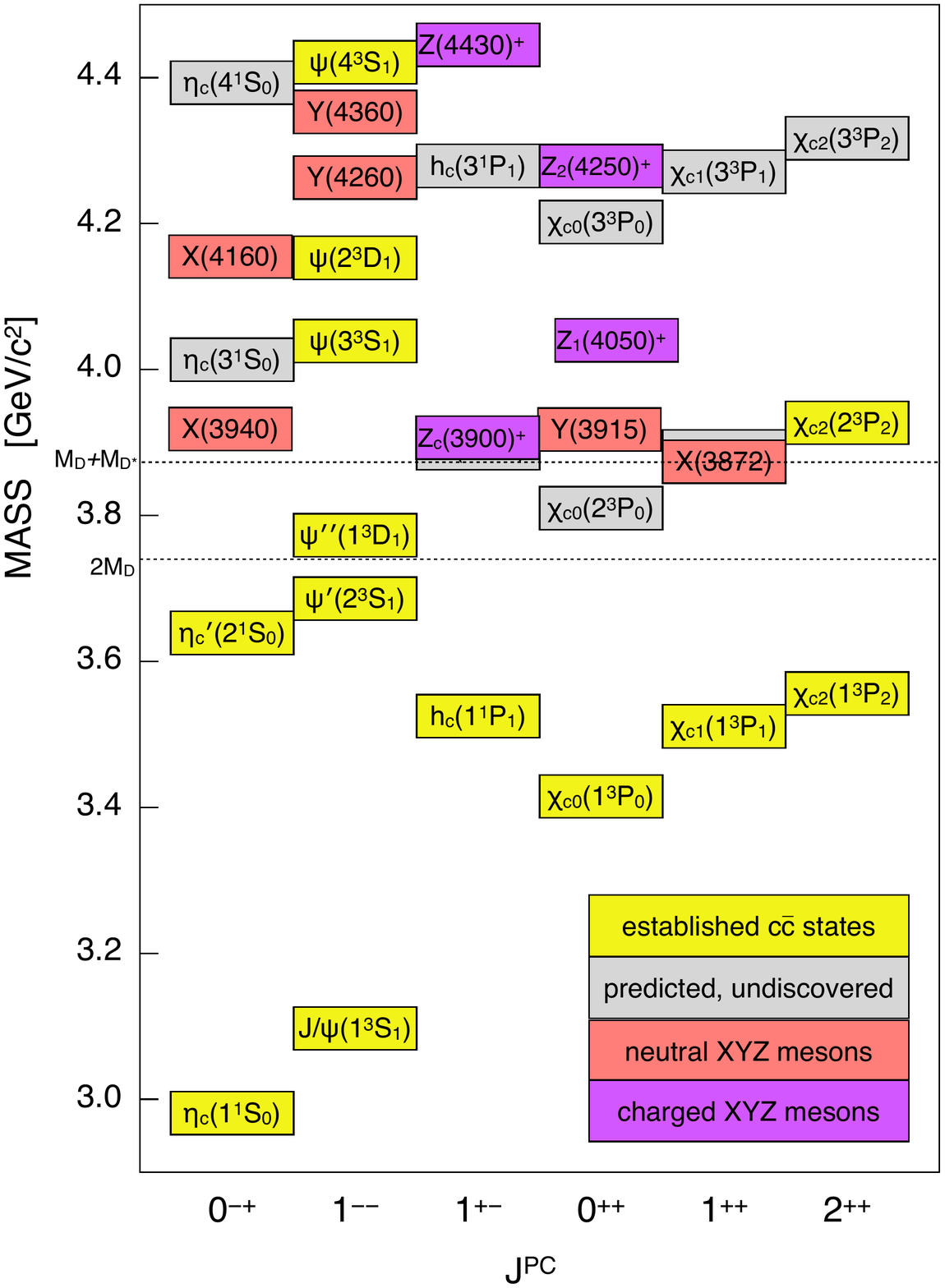}
\end{minipage}
\begin{minipage}[t]{85mm}
  \includegraphics[height=1.0\textwidth,width=1.0\textwidth]{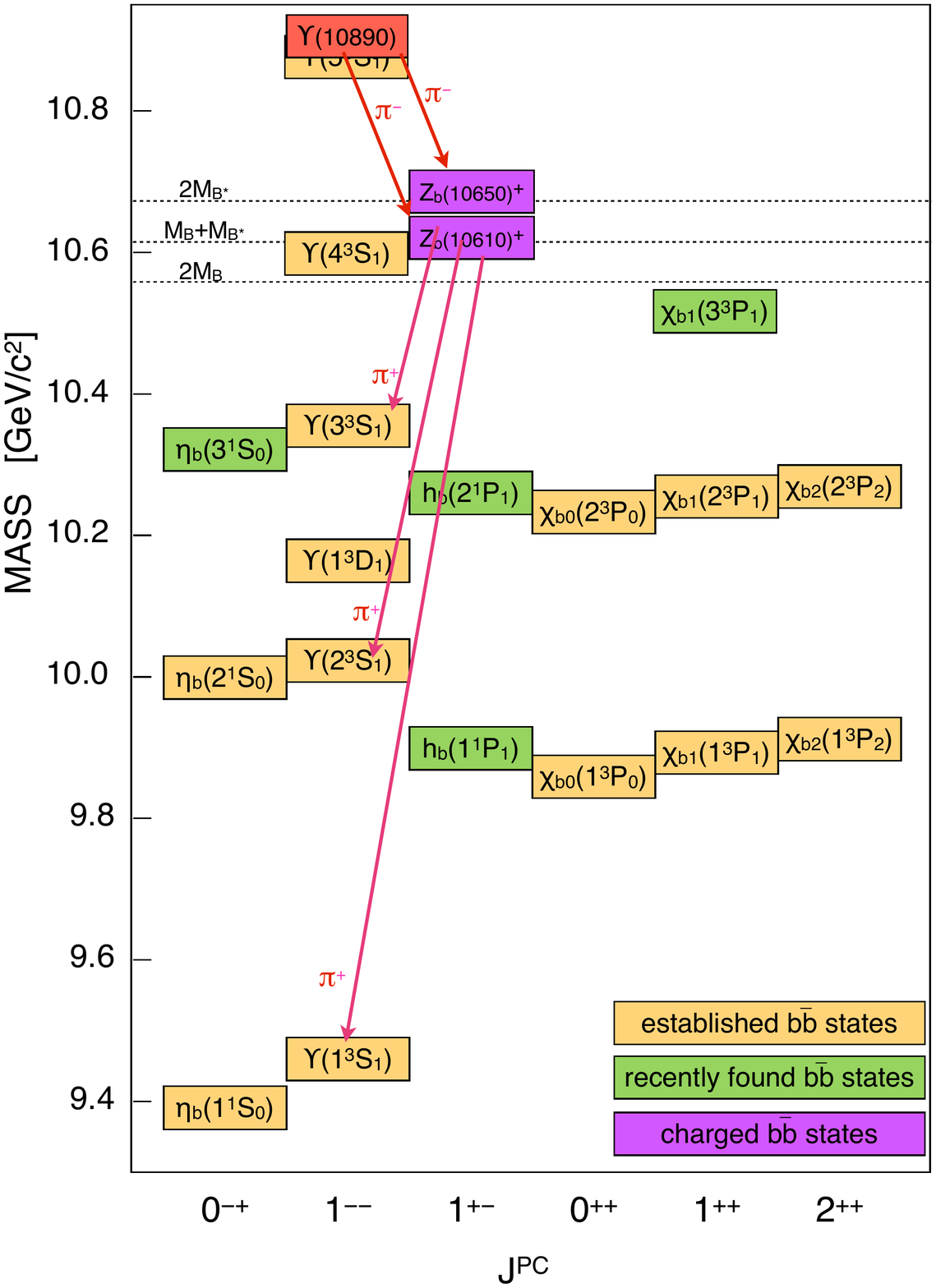}
\end{minipage}
\hspace{\fill}
\caption{\footnotesize The spectra of {\bf left)} charmoniumlike and
{\bf right)} bottomoniumlike mesons. 
}
\label{fig:xyz-levels}
\end{figure}

Figure~\ref{fig:xyz-levels} (left) shows the charmonium and charmoniumlike meson 
spectrum for masses below 4500~MeV. Here the yellow boxes indicate established charmonium states.
All of the
(narrow) states below the $2m_D$ open-charm threshold have been established and found to
have properties that are well described by the charmonium model.  In addition all of the
$J^{PC}=1^{--}$ states above $2m_D$ have also been identified.  The gray boxes show the
remaining unfilled, but predicted charmonium states.  The red boxes show electrically
neutral $X$ and $Y$ mesons and the purple boxes show the charged $Z$ mesons, aligned
according to my best guess at the $J^{PC}$ quantum numbers of their neutral charged partner.  
(Not included here are recently discovered resonances in the $\jp\phi$ channel, which are
discussed in a recent review by Yi~\cite{psiphi_kai}.)
In the following, I briefly
comment on each of the $XYZ$ entries.

\begin{figure}[htb]
%Figure with side by side by side panels
\begin{minipage}[t]{44mm}
  \includegraphics[height=0.7\textwidth,width=1.2\textwidth]{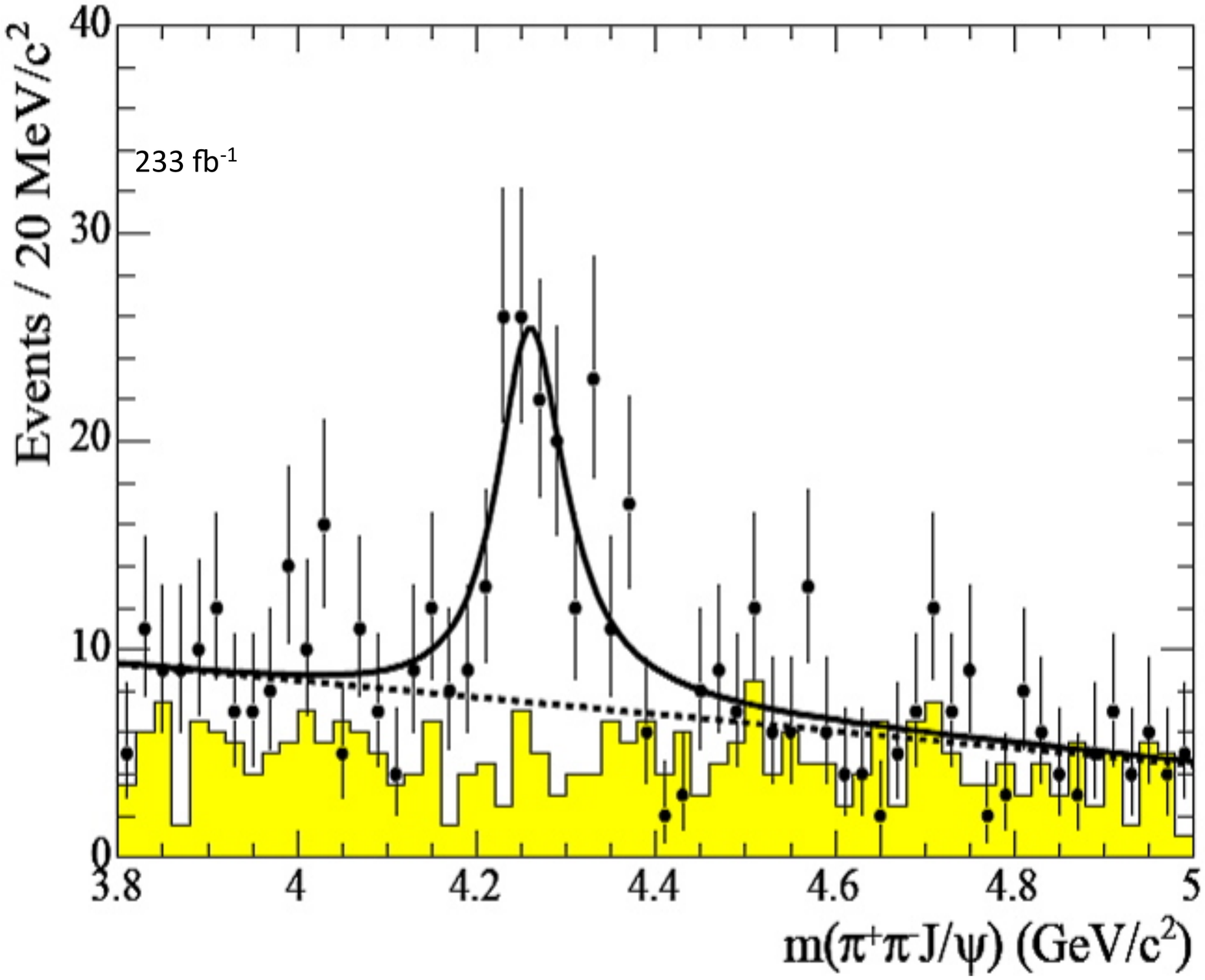}
\end{minipage}
\begin{minipage}[t]{40mm}
  \includegraphics[height=0.8\textwidth,width=1.4\textwidth]{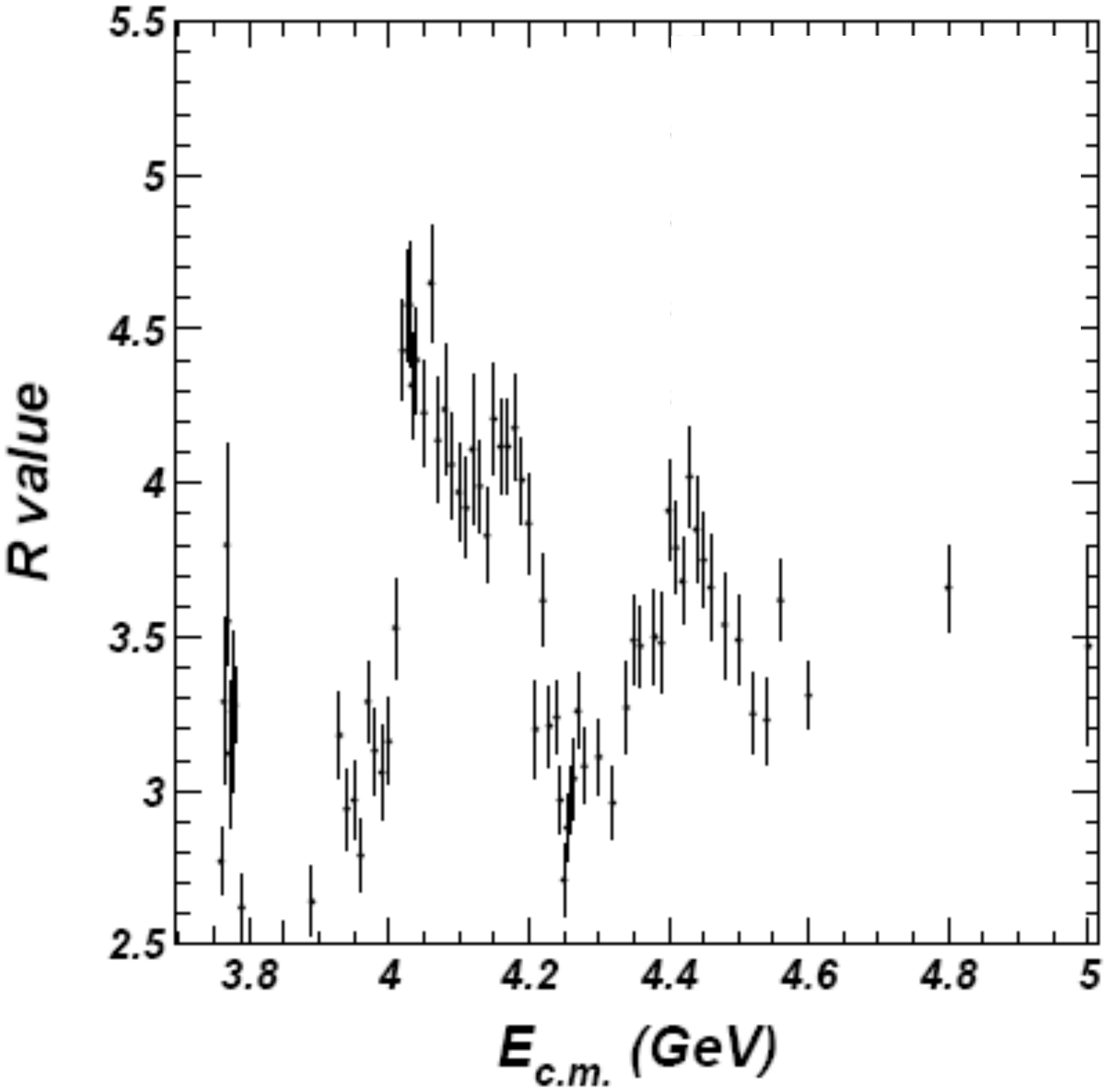}
\end{minipage}
\begin{minipage}[t]{40mm}
  \includegraphics[height=0.8\textwidth,width=1.2\textwidth]{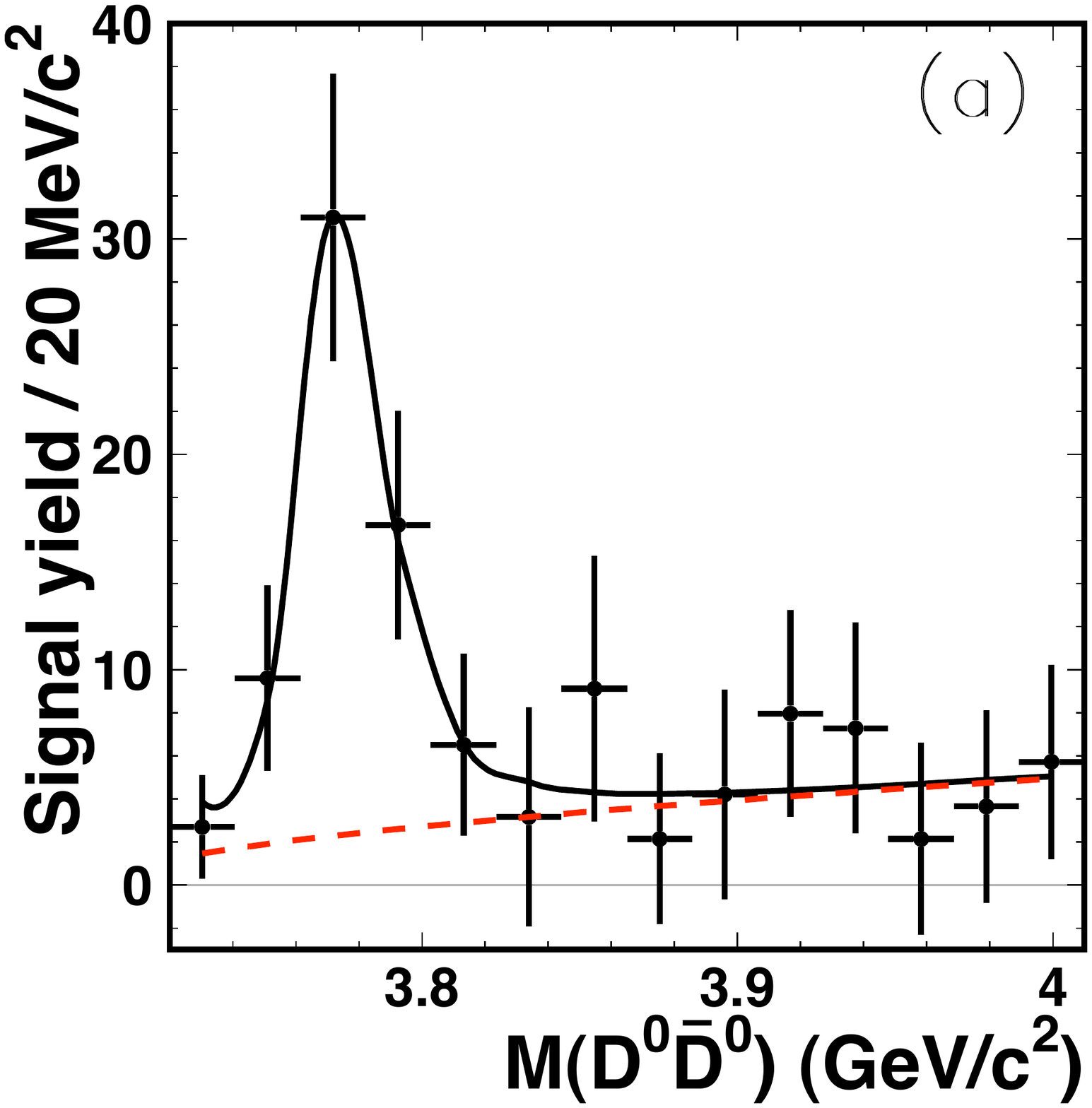}
\end{minipage}
\begin{minipage}[t]{44mm}
  \includegraphics[height=0.7\textwidth,width=1.2\textwidth]{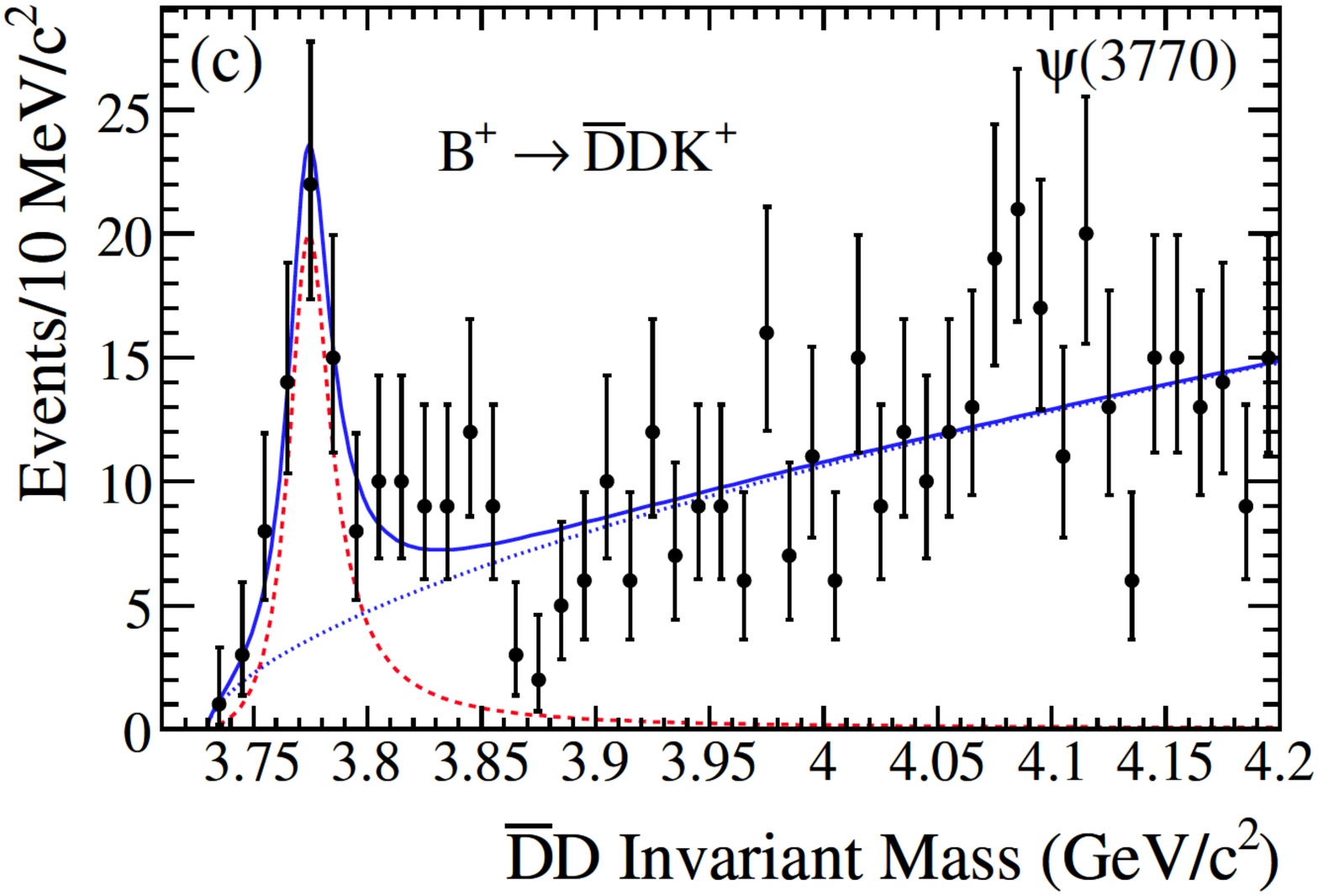}
\end{minipage}
\hspace{\fill}
\caption{\footnotesize {\bf Left)}  The $M(\pipi\jp)$ distribution
from $\ee\rt\gamma_{isr}\pipi\jp$ process near $\sqrt{s}=10.6$~GeV from ref.~\cite{babar_y4260}.
{\bf Center-left)} Inclusive Born cross section for $\ee$ annihilation into hadrons from
ref.~\cite{bes2_R}. 
{\bf Center-right)} The $M(D^0\bar{D}^0)$ distribution from $B\rt KD^0\bar{D}^0$ decays
from Belle~\cite{belle_KDD}.
{\bf Right)} The $M(D\bar{D})$ distribution from $B\rt KD\bar{D}$ decays from BaBar~\cite{babar_KDD}.
}
\label{fig:KDD}
\end{figure}

\begin{description}  

\item[\boldmath $X(3940)$ and $X(4160)$]~~The $X(3940)$ and $X(4160)$ were found by Belle in the
$D\bar{D}^*$~\cite{belle_x3940} and $D^*\bar{D}^*$~\cite{belle_x4160} systems recoiling against
a $\jp$ in $\ee\rt\jp D^{(*)}\bar{D}^*$ annihilations, respectively~\cite{conj}; their production mechanism
and their decay modes provide strong circumstantial evidence for $J^{PC}=0^{-+}$ assignments.
In both cases, the measured masses are far below expectations for the only available $0^{-+}$
charmonium levels, the $\eta_c(3S)$ and $\eta_c(4S)$.  These assignments would imply
hyperfine $n^3S-n^1S$ mass splittings that increase from the measured value of $47.2\pm 1.2$~MeV
for $n=2$~\cite{PDG}, to $\sim 100$~MeV for $n=3$ and $\sim 350$~MeV for $n=4$.  This is problematic
because potential models predict that hyperfine splittings decrease with increasing $n$.

\item[\boldmath $Y(4260)$ and $Y(4360)$]~~BaBar discovered the $Y(4260)$ as a peak in the $\pipi\jp$
system produced in the initial-state-radiation process $\ee\rt\gamma_{isr}\pipi\jp$~\cite{babar_y4260}
(see the left panel of Fig.~\ref{fig:KDD}) and the $Y(4360)$ in the $\pipi\psi^{\prime}$ system produced
via $\ee\rt\gamma_{isr}\pipi\psi^{\prime}$~\cite{babar_y4360}.  These states are considered to be exotic since the
production mechanism ensures that their $J^{PC}=1^{--}$ while all of the $1^{--}$ $\ccbar$ states near
their masses have already been assigned.  Moreover, there is no evidence for them
in any exclusive~\cite{galina} or the inclusive~\cite{bes2_R} charmed meson pair production cross
section, where there is a pronounced dip at $\sqrt{s}\simeq 4.26$~GeV (see the center-left panel of Fig.~\ref{fig:KDD}).
This dip implies a large partial width for $Y(4260)$ decay 
to $\pipi\jp$: $\Gamma(Y(4260)\rt \pipi\jp)>1$~MeV~\cite{mo}, which is huge by charmonium
standards.

\item[\boldmath $Z_c(3900)$, $Z_1(4050)$, $Z_2(4250)$ and $Z(4430)$]~~Since the
$Z_c(3900)$~\cite{bes_z3900}, $Z_1(4050)$~\cite{belle_z1z2}, $Z_2(4250)$~\cite{belle_z1z2}
and $Z(4430)$~\cite{belle_z4430} are electrically charged and decay to hidden charm final states,
their minimal quark structure must be a $\ccbar u\bar{d}$ four-quark combination and they are, therefore,
exotic.

\item[\boldmath $Y(3915)$]~~The $Y(3915)$ is an $\omega\jp$ mass peak at $3918\pm 3$~MeV seen
in $B\rt K\omega\jp$ decays~\cite{belle_y3940,babar_y3940} and in the two-photon process
$\gamma\gamma\rt\omega\jp$~\cite{belle_2g_y3915,babar_2g_y3915}.  BaBar measured its $J^{PC}$
quantum numbers to be $0^{++}$~\cite{babar_jpc_y3915}.  The PDG currently assigns this as the 
$\chi_{c0}(2P)$ charmonium level, an assignment that has two problems: {\it i}) since the $\chi_{c2}(2P)$
has been established with mass $3927\pm 3$~MeV, the $Y(3915)=\chi_{c0}(2P)$ assignment would imply that the
$2^3P_2 - 2^3P_0$ fine splitting is only $9\pm 4$~MeV, and tiny in comparison with the corresponding
$n=1$ splitting of $141.4\pm 0.3$~MeV~\cite{PDG}; {\it ii}) there is no sign of $Y(3915)\rt D\bar{D}$
decay. Belle~\cite{belle_KDD} and BaBar~\cite{babar_KDD} have studied the process $B\rt K D\bar{D}$,
and both groups see a prominent signal for $\psi(3770)\rt D\bar{D}$ but no hint of a $D\bar{D}$ mass
peak near 3915~MeV (see the two right-most panels in Fig.~\ref{fig:KDD}).  Since neither
group reported any $Y(3915)$-related limit, I derived my own conservative upper limit by
scaling the total number of Belle events in the two mass bins surrounding 3915~MeV in the
center-right panel of Fig.~\ref{fig:KDD} to the $\psi(3770)$ signal, while assuming constant acceptance.
This gives ${\mathcal B}(Y(3915)\rt D\bar{D})<{\mathcal B}(Y(3915)\rt \omega\jp),$ which strongly
contradicts theoretical expectations that for the $\chi_{c0}(2P)$, the $D\bar{D}$ decay channel should
dominate~\cite{barnes}.
(Note: some of the literature -- and the PDG -- refer to this state as the $X(3915)$.)
  
\item[\boldmath $X(3872)$]~~The $X(3872)$ was first seen in 2003 as a peak in the $\pipi\jp$
invariant mass distribution in $B\rt K\pipi\jp$ decays~\cite{belle_x3872}.  Its mass is
indistinguishable from the $m_{D^0} + m_{D^{*0}}$ threshold; the most recent result from LHCb
is $M_{X(3872)}-(m_{D^0}+m_{D^{*0}}) = -0.09\pm 0.28$~MeV~\cite{eidelman}.  LHCb also established
the $J^{PC}$ quantum numbers as $1^{++}$~\cite{lhcb_jpc}.   The $X(3872)$ has a long interesting story that
I will not attempt to even summarize in this brief report.  Instead I refer the reader to
ref.~\cite{kkseth} and references cited therein.

\end{description}

\subsubsection{Bottomoniumlike mesons}

\begin{figure}[htb]
%Figure with side by side by side panels
\begin{minipage}[t]{85mm}
  \includegraphics[height=1.0\textwidth,width=1.0\textwidth]{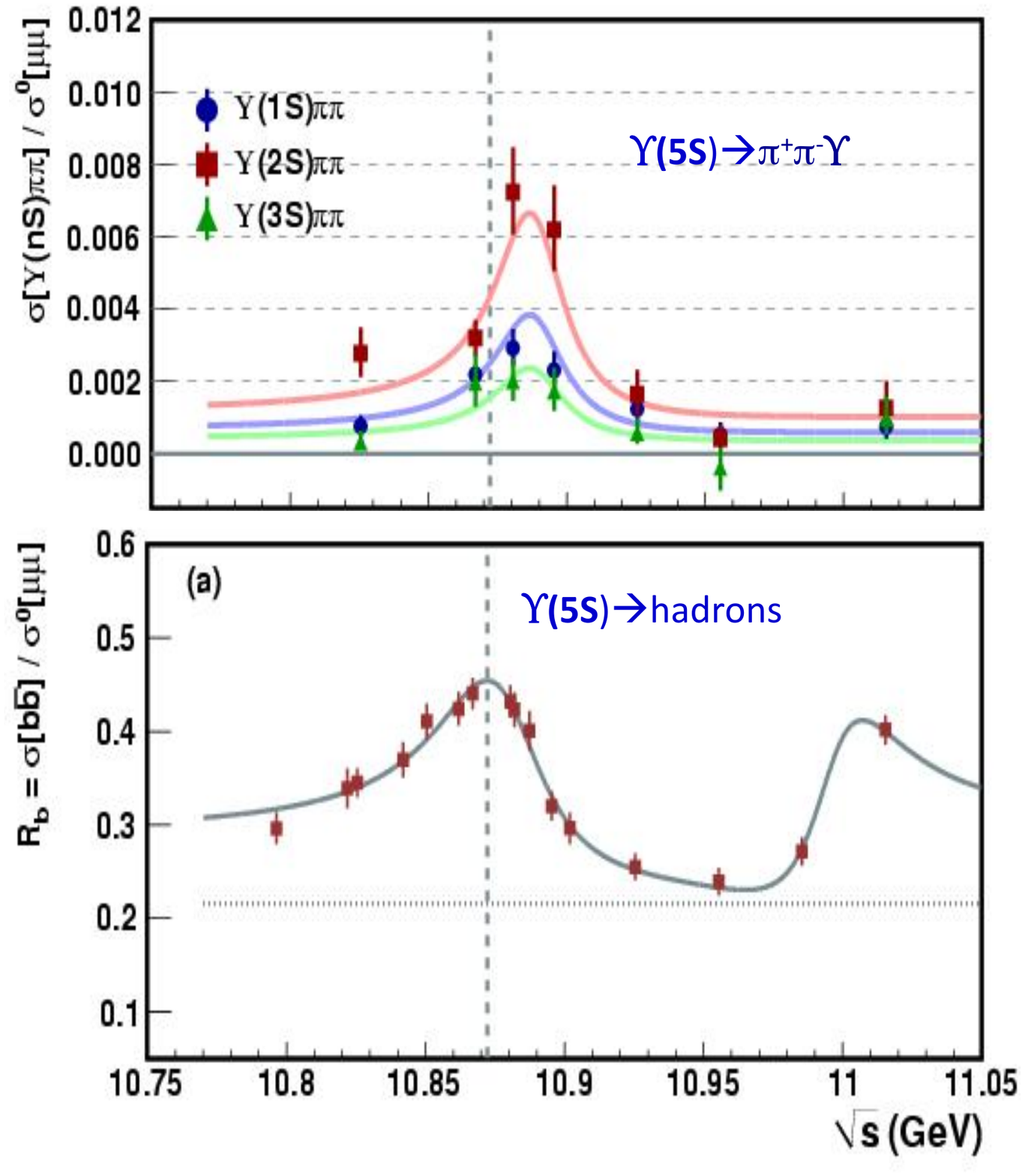}
\end{minipage}
\begin{minipage}[t]{85mm}
  \includegraphics[height=1.0\textwidth,width=1.0\textwidth]{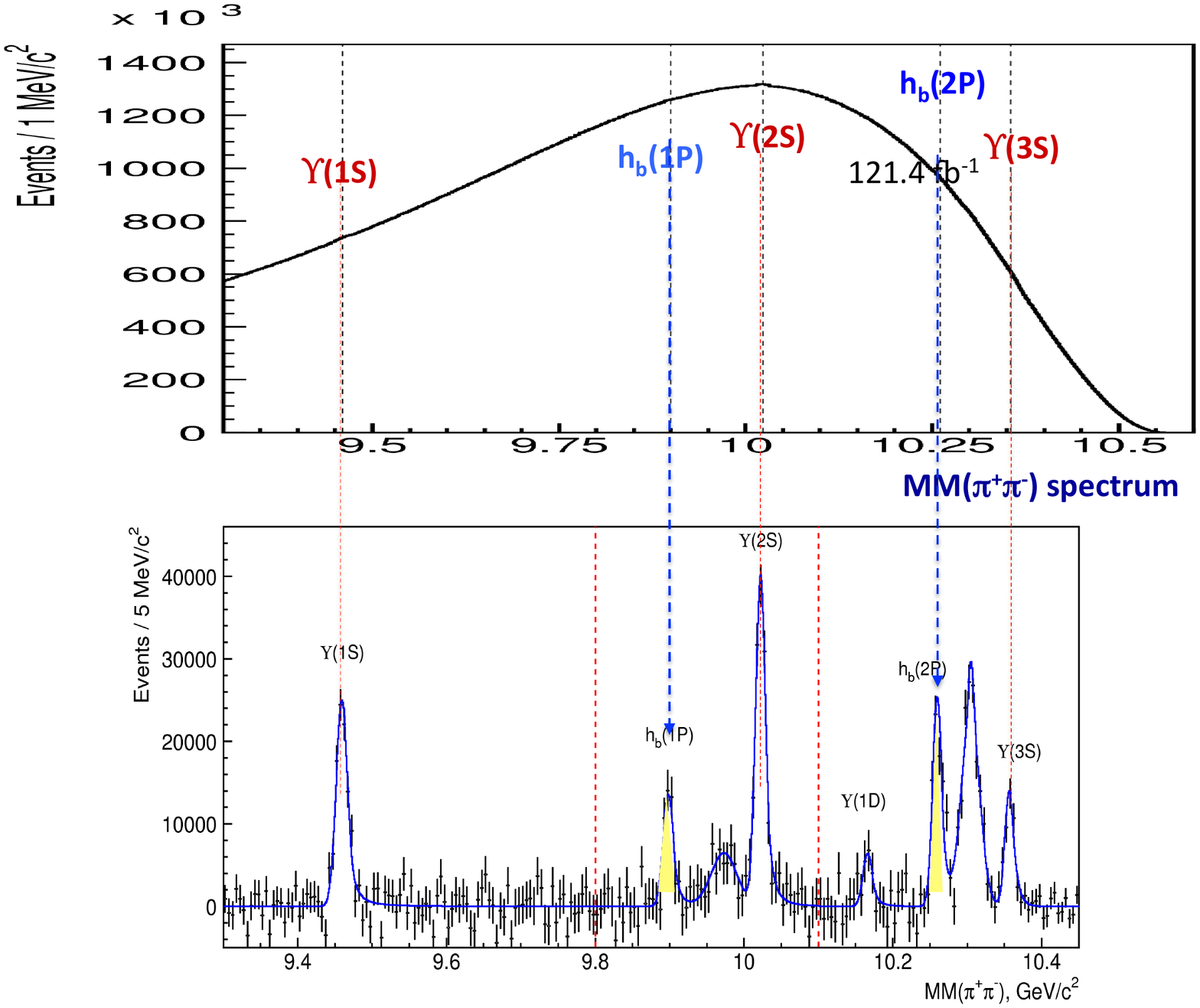}
\end{minipage}
\hspace{\fill}
\caption{\footnotesize {\bf Left)} Cross sections for $\ee\rt \pipi\Upsilon (nS)$
($n=1,2,3$) {\it (upper)} and $\ee\rt~hadrons$ ({\it (lower)} in the vicinity of the
$\Upsilon(5S)$ resonance from ref.~\cite{belle_kfchen2}. {\bf Right)} Distribution of
masses recoiling against $\pipi$ pairs at $\sqrt{s}\simeq 10.87$~GeV {\it (upper)}
and residuals from piecewise fits to the data with smooth polynomials {\it (lower)} 
(from ref.~\cite{belle_hb}).  The $h_b(1P)$ and $h_b(2P)$ peaks are shaded in yellow.
 }
\label{fig:ups5s}
\end{figure}

The right panel of Fig.~\ref{fig:xyz-levels} shows the recent status of the $\bbbar$ bottomonium
and bottomoniumlike mesons. Here the orange boxes indicate the well established bottonium mesons and
the green boxes show those that were recently established.  The large $Y(4260)\rt\pipi\jp$ signal seen in the
charmonium mass region motivated a Belle search for similar behavior in the bottomonium system~\cite{wshou}
that uncovered an anomalously large $\pipi\Upsilon(nS)$ $(n=1,2,3)$ production rates that peak around
$\sqrt{s}=10.89$~GeV as shown in the upper-left panel of Fig.~\ref{fig:ups5s}~\cite{belle_kfchen2}.  The peak
position of the $\pipi\Upsilon(nS)$ yield is about $2\sigma$ higher than that of the peak in the $\ee\rt~hadron$
cross section at $\sqrt{s}\simeq10.87$~GeV that is usually associated with the conventional $\Upsilon(5S)$ $\bbbar$
meson (and shown in the lower left panel of Fig.~\ref{fig:ups5s}).  If the peak in the $\pipi\Upsilon(nS)$
cross section is attributed to the $\Upsilon(5S)$, it implies  $\Upsilon(5S)\rt \pipi\Upsilon(nS)$, ($n=1,2,3$) 
partial widths that are hundreds of times larger than both theoretical predictions~\cite{belle_kfchen1} and the
corresponding measured values for the $\Upsilon(4S)$~\cite{PDG}).   This suggests the presence of a $\bbbar$
equivalent of the $Y(4260)$ with mass near $10,890$~MeV, as suggested in ref.~\cite{wshou}. 

The upper right panel of Fig.~\ref{fig:ups5s} shows the distribution of
masses recoiling against all of the $\pipi$ pairs in events collected at
near $\sqrt{s}\simeq 10.87$~GeV, near the peak of the $\Upsilon(5S)$
resonance.  The combinatoric background is huge -- there are typically
$10^6$ entries in each 1~MeV bin -- and the statistical errors are $\sim 0.1\%$. 
The data were fit piece-wise with sixth-order
polynomials, and the residuals from the fits are shown in the lower right
panel of Fig.~\ref{fig:ups5s}, where, in addition to peaks
at the $\yones$, $\ytwos$, $\Upsilon(3S)$ and some expected reflections,
there are unambiguous signals for the $h_b(1P)$ and $h_b(2P)$, the $1^1P_1$
and $2^1P_1$ bottomonium states.  This was the first observation of these two
elusive levels~\cite{belle_hb}.  One puzzle is that the $\pipi h_b(mS)$, ($m=1,2$) 
final states are produced at rates that are nearly the same as those for 
$\pipi\Upsilon(nS)$, ($n=1,2,3$), even though the $\pipi h_b$ transition requires a
heavy-quark spin flip, which should result in a strong suppression.  This motivated a more
detailed investigation of these channels.

\begin{figure}[htb]
%Figure with side by side by side panels
\begin{minipage}[t]{95mm}
  \includegraphics[height=1.0\textwidth,width=1.0\textwidth]{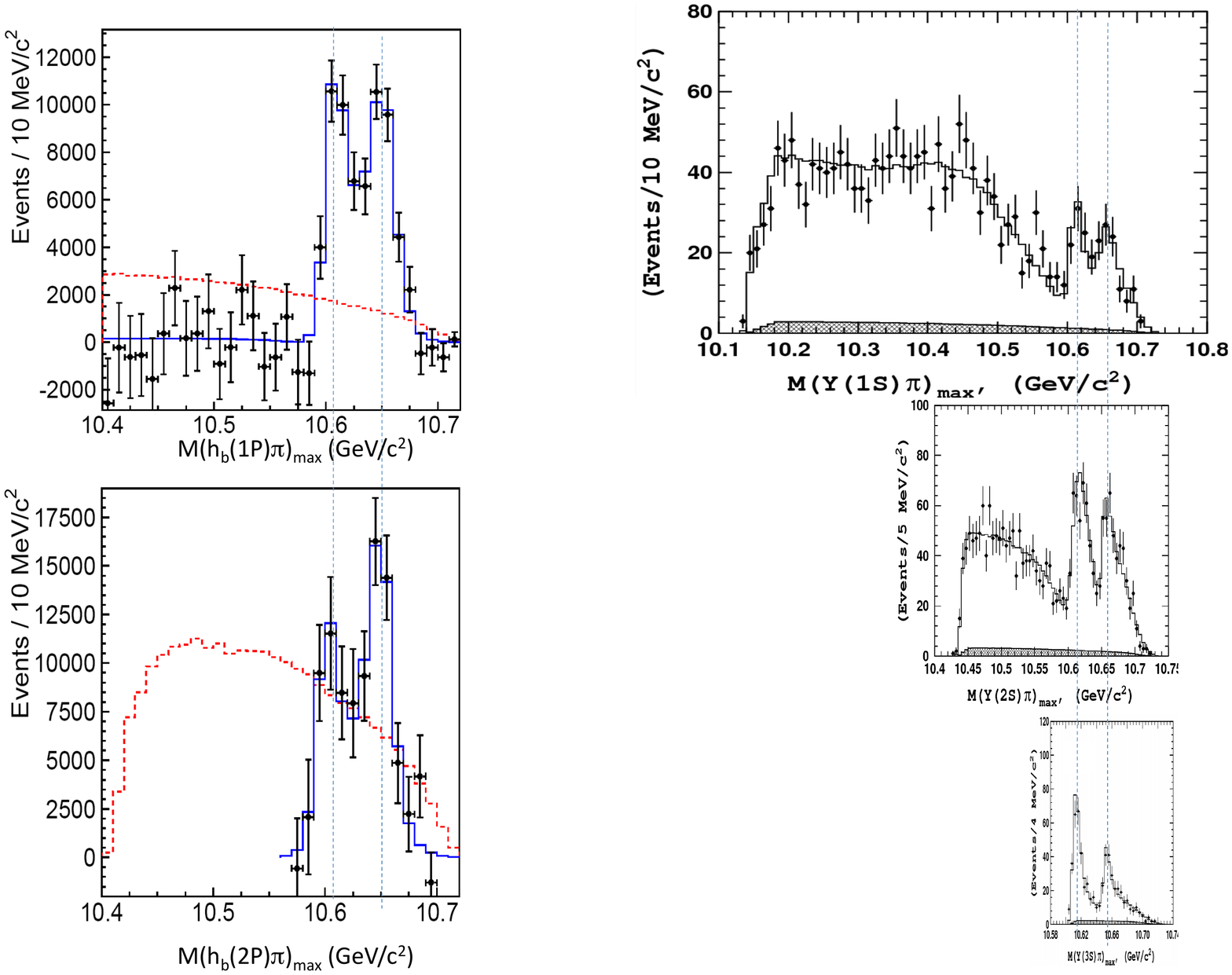}
\end{minipage}
\begin{minipage}[t]{75mm}
  \includegraphics[height=1.25\textwidth,width=1.4\textwidth]{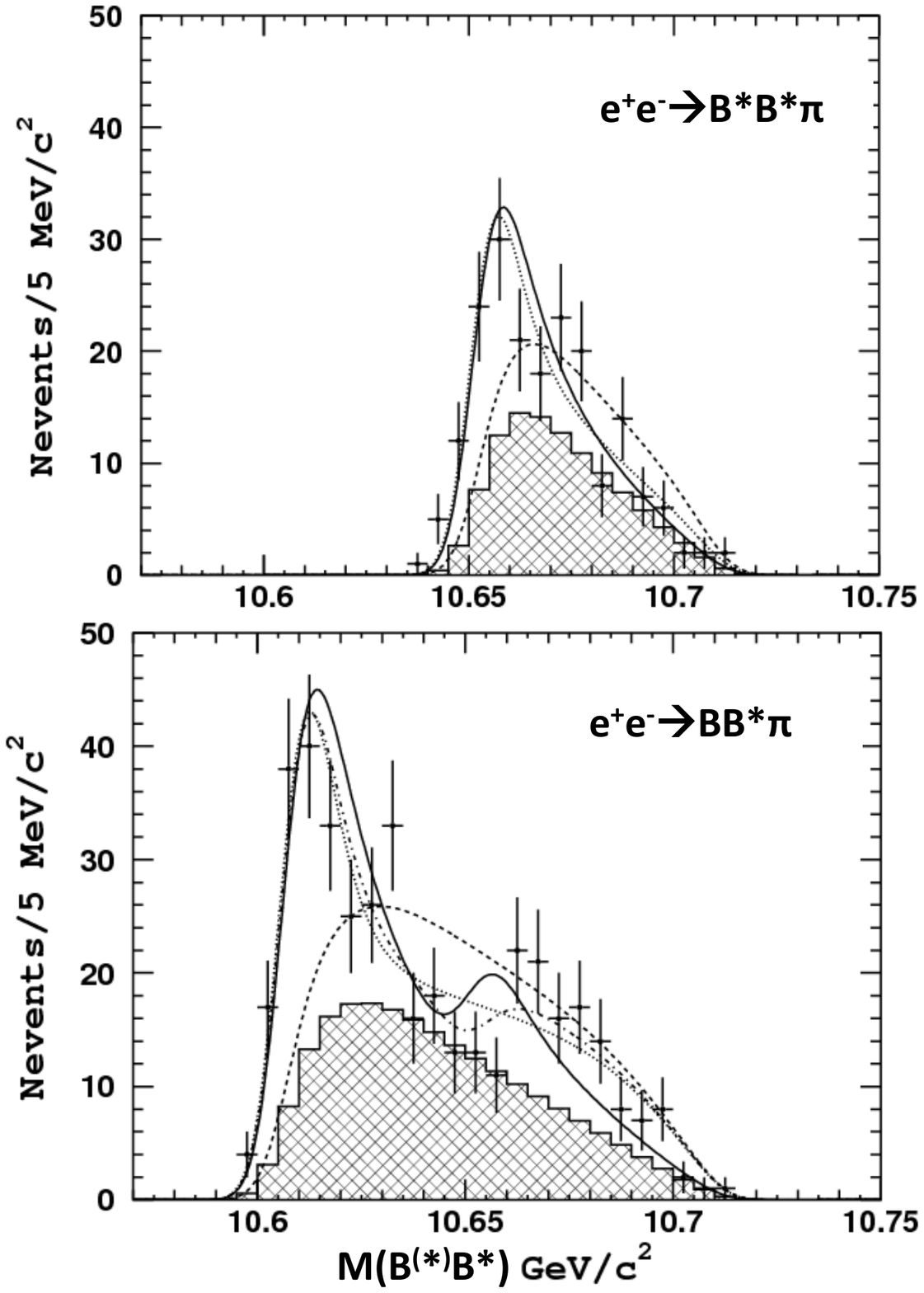}
\end{minipage}
\hspace{\fill}
\caption{\footnotesize {\bf Left)} invariant mass distributions for $h_b(1P)\pip$
{\it (upper)} and $h_b(2P)\pip$ {\it (lower)} from $\ee\rt\pipi h_b(nP)$ events.
{\bf Center)} Invariant mass distributions for $\yones\pip$ {\it (upper)}, $\ytwos\pip$
{\it (center)} and $\Upsilon(3S)\pip$ {\it (lower)} in $\ee\rt\pipi\Upsilon(nS)$ events.
the figures are from ref.~\cite{belle_zb}, and scaled to make the make the horizontal
scales (almost) match. 
{\bf Right)} The $M(B^*\bar{B}^*)$ {\it (upper)} and $M(B\bar{B}^*)$ {\it (lower)} invariant mas distributions
from $\ee\rt B^{(*)}\bar{B}^*\pi$ events near $\sqrt{s}=10.97$~GeV (from ref.~\cite{belle_bbstr}).
}
\label{fig:zb}
\end{figure}

The left panels of Fig.~\ref{fig:zb} show the $\pipi h_b$ yield versus the
maximum $h_b \pi^{\pm}$ invariant mass for $h_b=h_b(1P)$ (upper) and $h_b=h_b(2P)$ (lower),
where it can be seen that all of the $\pipi h_b$ events are concentrated in two
$M_{\rm max}(h_b\pi)$ peaks, one near $10,610$~MeV and the other near $10,650$~MeV~\cite{belle_zb}.  Studies
of fully reconstructed $\pipi \Upsilon(nS)$, ($n=1,2,3)$, $\Upsilon(nS)\rt\ell^+\ell^-$
events in the same data sample show two peaks in the $M_{\rm max}(\Upsilon(nS)\pi)$ 
distributions at the same masses for all three
modes, as shown in the three center panels of Fig.~\ref{fig:zb}. Here the fractions of $\pipi\Upsilon(nS)$
events in the two peaks are substantial -- $\sim 6\%$ for the $\yones $, $\sim 22\%$ for the $\ytwos$
and $\sim 43\%$ for the $\Upsilon(3S)$ -- but do not account for all of the $\pipi\Upsilon(nS)$ event yield~\cite{bondar}.
The peak masses and widths in all five channels are consistent with each other and the
weighted average mass and width values are $M_{Z_b(10610)}=10607.2\pm 2.0$~MeV \&
$\Gamma_{Z_b(10610)}=18.4\pm 2.4$~MeV, and $M_{Z_b(10650)}=10652.2\pm 1.5$~MeV \&
$\Gamma_{Z_b(10650)}=11.5\pm 2.2$~MeV.  Dalitz-plot analyses of the $\pipi\Upsilon(nS)$
final states establish $J^P=1^+$ quantum number assignments for both states.
The lower mass state is just $2.6\pm 2.2$~MeV above the $m_B+m_{B^*}$ mass threshold and the higher mass state
is only $2.0\pm 1.6$~MeV above $2m_{B^*}$, which, together with the $J^P=1^+$ assignment,
is suggestive of virtual $B\bar{B}^*$ and $B^*\bar{B}^*$ $S$-wave molecule-like states.

The $B^{(*)}\bar{B}^{*}$ molecule picture is supported by a Belle study of $\ee\rt B^{(*)}\bar{B}^*\pi$
final states in the same data sample~\cite{belle_bbstr}, where $B\bar{B}^*$ and $B^*\bar{B}^*$
invariant mass peaks are seen at the $Z_b(10610)$ and $Z_b(10650)$ mass values, respectively,
as shown in the right panels of Fig.~\ref{fig:zb}.  From these data, preliminary values
of the branching fractions 
${\mathcal B}(Z_b(10610)^+\rt B^+\bar{B}^{*0} + \bar{B}^0B^{*+} + c.c.) = (86.0\pm 3.6)\%  $ and 
${\mathcal B}(Z_b(10610)^+\rt B^{*+}\bar{B}^{*0} + c.c.) = (73.4\pm 7.0)\%  $ are inferred.
The measured branching fraction for $Z_b(10610)\rt B^{*}\bar{B}^{*}$ is consistent with zero.
This pattern, where $B\bar{B}^*$ decays dominate for the $Z_b(10610)$ and $B^*\bar{B}^*$
decays are dominant for the $Z_b(10650)$ are consistent with expectations for
molecule-like structures.

\subsubsection{Operating BESIII/BEPCII as a $Y(4260)$ factory}

As described above, 
the peculiar properties of the $Y(4260)$ charmoniumlike state motivated
the Belle studies of the $\pipi\Upsilon(nS)$ and $\pipi h_b(mP)$ channels at energies near the
peak of the $\Upsilon(5S)$ and led to the discovery of the charged, bottomoniumlike $Z_b$
mesons.  These discoveries, in turn, inspired the BESIII group to revisit the $Y(4260)$ by operating the 
Beijing Electron-Positron Collider (BEPCII) at $\sqrt{s}=4.26$~GeV, and accumulating a large sample
of $Y(4260)$ decay events.  

\begin{figure}[htb]
%Figure with side by side by side panels
\begin{minipage}[t]{44mm}
  \includegraphics[height=1.0\textwidth,width=1.0\textwidth]{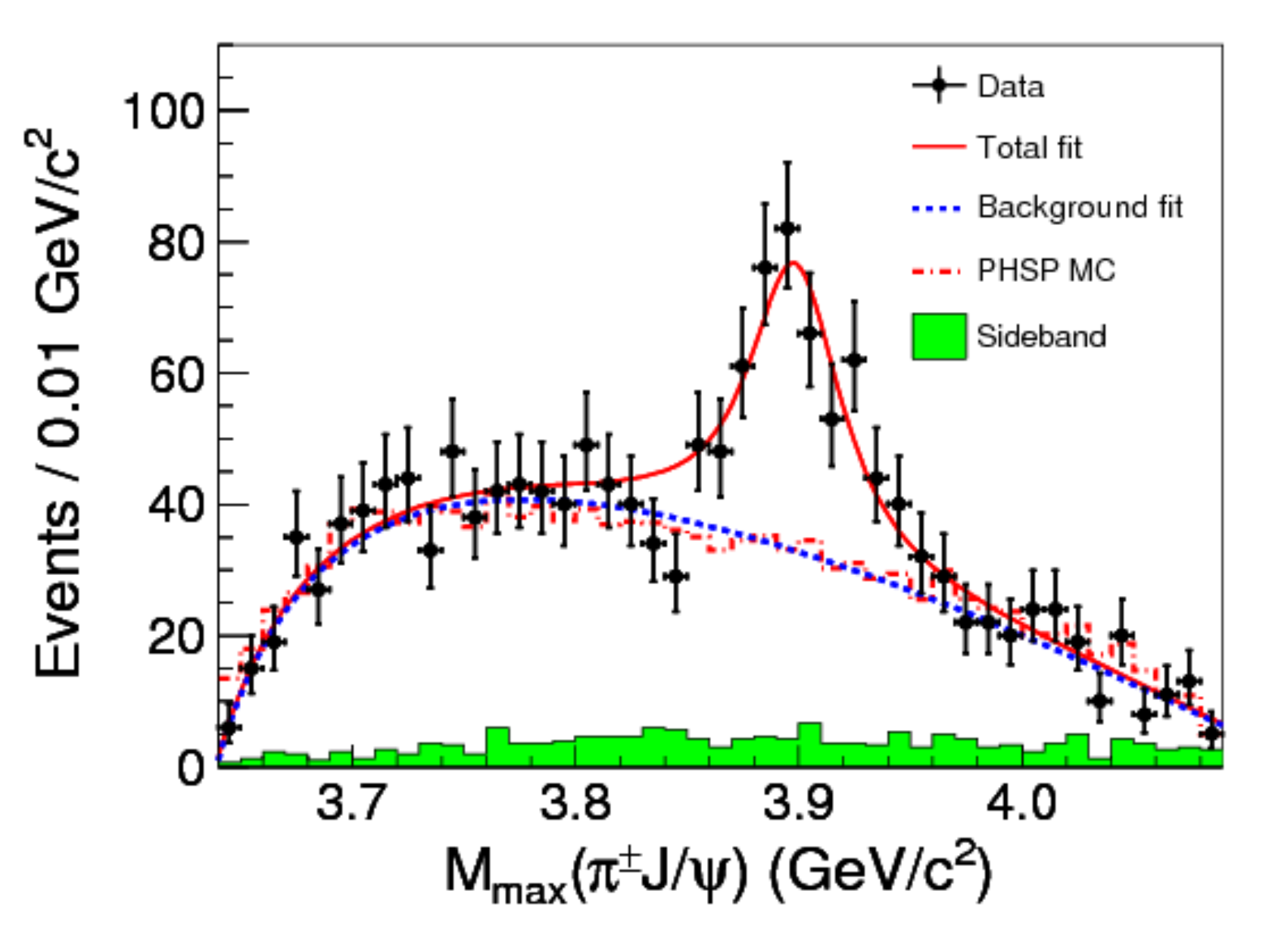}
\end{minipage}
\begin{minipage}[t]{44mm}
  \includegraphics[height=1.0\textwidth,width=1.0\textwidth]{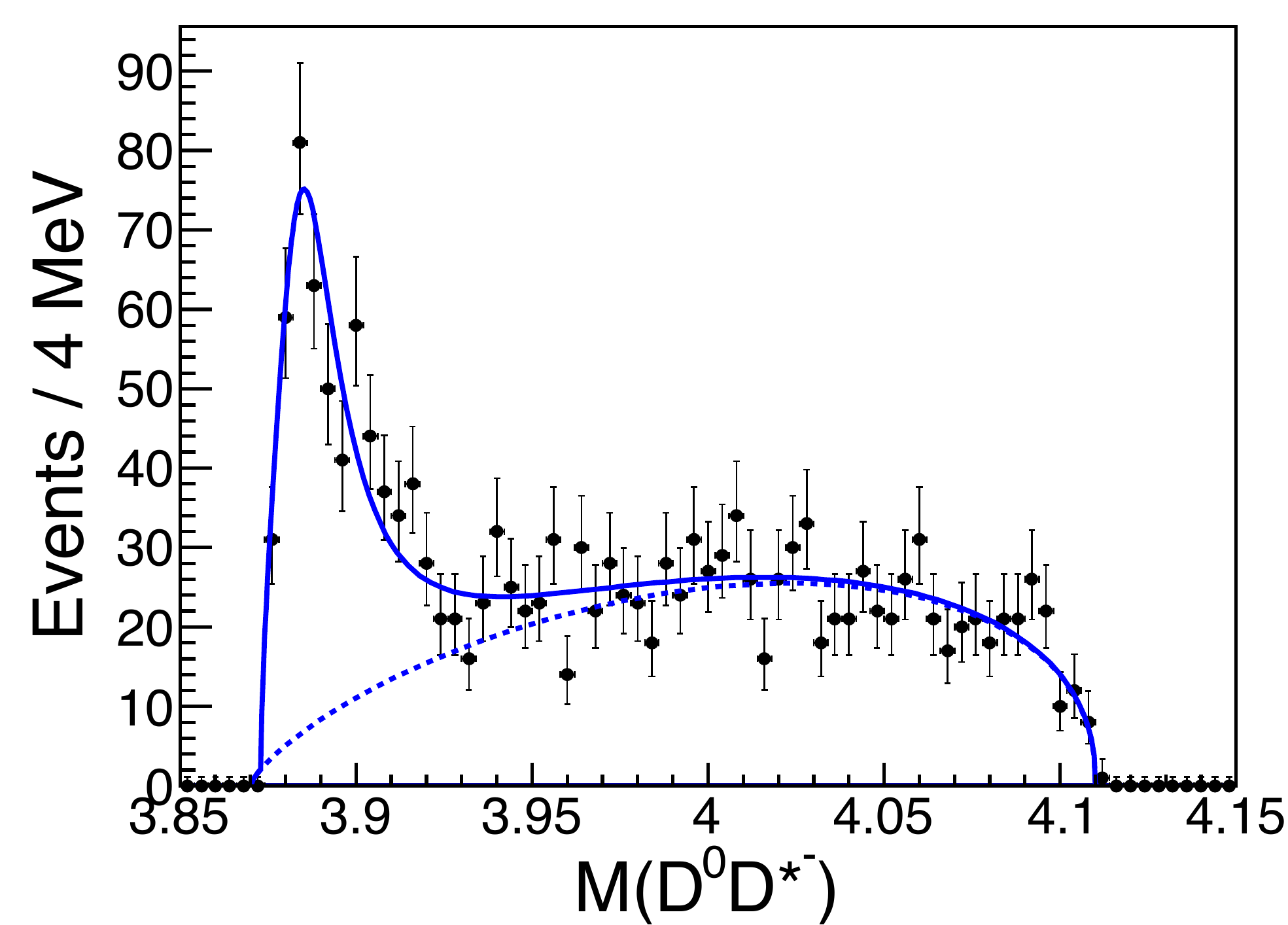}
\end{minipage}
\begin{minipage}[t]{44mm}
  \includegraphics[height=1.0\textwidth,width=1.0\textwidth]{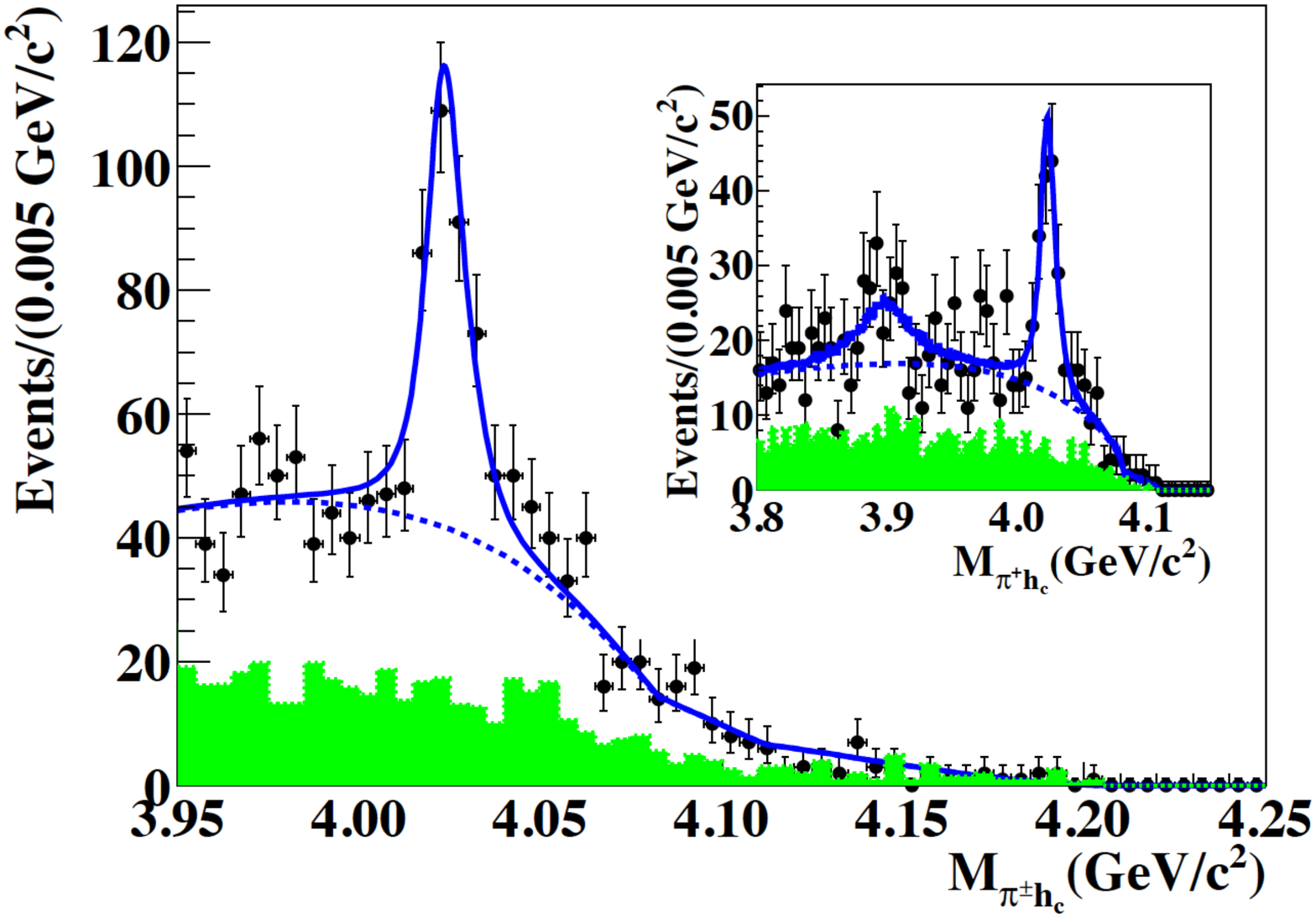}
\end{minipage}
\begin{minipage}[t]{44mm}
  \includegraphics[height=1.0\textwidth,width=1.0\textwidth]{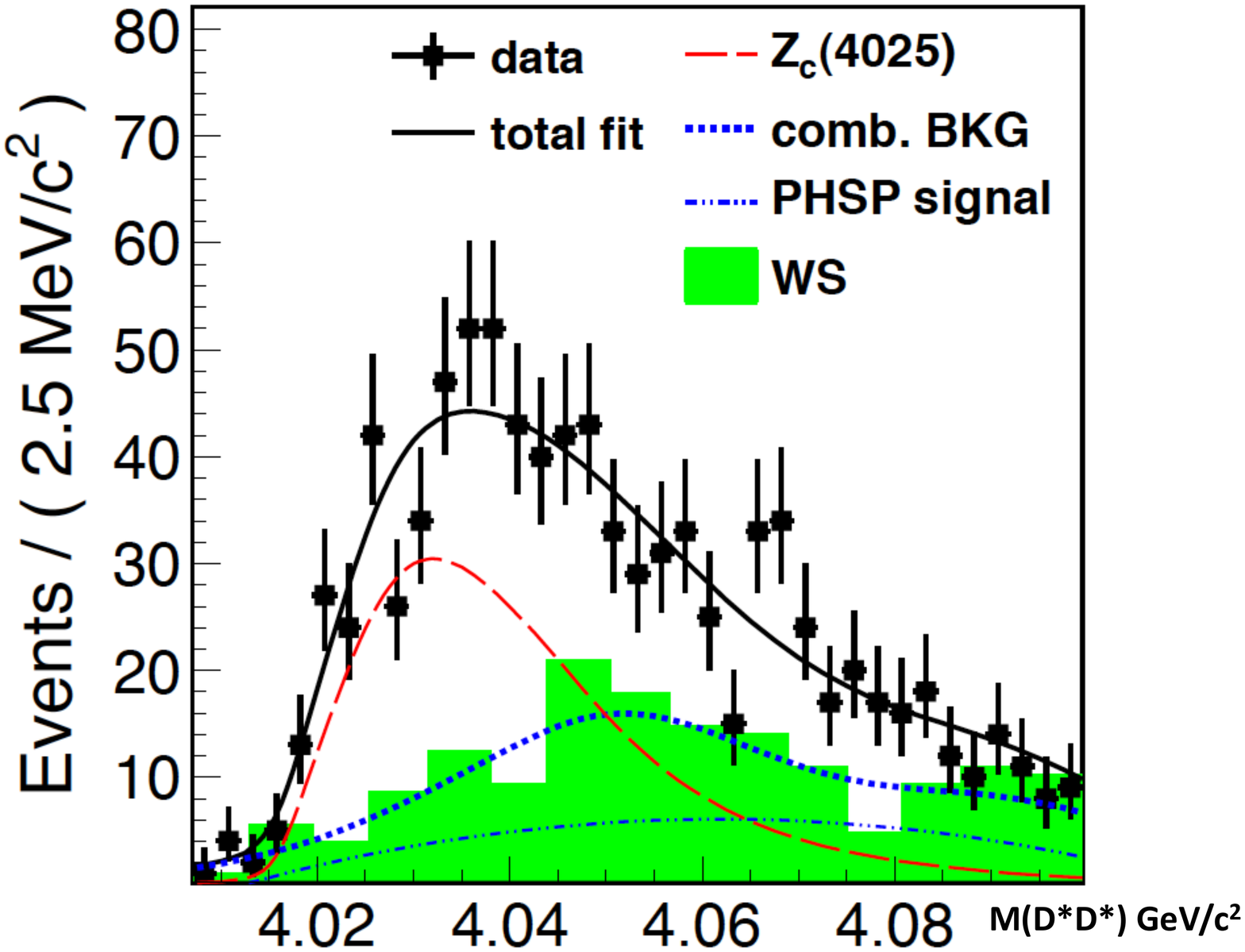}
\end{minipage}
\hspace{\fill}
\caption{\footnotesize Invariant mass distributions for $\jp\pip$ from $\ee\rt\pipi\jp$ events {\bf (Left)},
$D\bar{D}^*$ from $\ee\rt (D\bar{D}^*)^+\pim$ events {\bf (Center-left)}, $h_c(1p)\pip$
from $\ee\rt\pipi h_c(1P)$ events {\bf (Center-right)} near $\sqrt{s}=4.26$~GeV and
$D^*\bar{D}^*$ from $\ee\rt (D^*\bar{D}^*)^+\pim$ events {\bf (Right)} at $\sqrt{s}=4.26$~GeV.
}
\label{fig:zc}
\end{figure}

The first channel to be studied with these data was the $\ee\rt\pipi\jp$, where a distinct peak in the distribution of 
the larger of the two $\jp\pi^{\pm}$ invariant mass combinations in each event ($M_{\rm max}(\jp\pi)$), exhibits a distinct
peak near $M_{\rm max}(\jp\pi)\simeq 3900$~MeV, which is called the $Z_c(3900)$ and shown in the left panel of
Fig.~\ref{fig:zc}~\cite{bes_z3900}.
A fit using a mass-independent-width BW function to represent the peak yielded a mass and width of
$M_{Z_c(3900)}=3899.0\pm 6.1$~MeV and $\Gamma_{Z_c(3900)}= 46 \pm 22$ ~MeV, which is $\sim 20$~MeV above the $m_D + m_{D^*}$
threshold.   The $Z_c(3900)$, which looks like a charmed-sector version of the $Z_b(10610)$, was subsequently confirmed by
Belle~\cite{belle_z3900}.

The close proximity of the $Z_c(3900)$ mass to the $m_{D}+m_{D^*}$ threshold motivated a BESIII study of the $(D\bar{D}^*)^+$
systems produced in $(D\bar{D}^*)^{\pm}\pi^{\mp}$  final states in the same data sample~\cite{bes_z3885}.  
There, dramatic near-threshold peaks are seen in both the $D^0D^{*-}$ and $D^+\bar{D}^{*0}$  invariant mass distributions;
as an example, the $M(D^0D^{*-})$ distribution is shown in
the center-left panel of Fig.~\ref{fig:zc}.  The weighted average mass and width from fits of a threshold-modified BW
line shapes to these distributions gives resonance pole position ($M_{\rm pole} + i\Gamma_{\rm pole}$) values of
$M_{\rm pole}= 3883.9\pm 4.5$~MeV and $\Gamma_{\rm pole}=24.8\pm 12$~MeV.  For these data, the production angle dependence
strongly favors a $J^P=1^+$ quantum number assignment and decisively rules out $J^P=0^-$ and $1^-$ ($0^+$ is ruled out by
parity conservation).  Since the pole mass position
is $\simeq 2\sigma$ lower than the $Z_c(3900)$ mass, BESIII tentatively named this $D\bar{D}^*$ state the $Z_c(3885)$.
In the mass determinations of both the $Z_c(3885)$ and $Z_c(3900)$, effects of possible interference with a coherent 
component of the background are ignored. Since this can bias the measurements by amounts comparable to the resonance widths,
which are large, this may account for the discrepancy in masses. A partial wave analysis of the $\pipi\jp$ channel, which
is currently underway, may establish the $J^P$ quantum numbers of
the $Z_c(3900)$ and provide important input into the question of whether or not these are the same state.  If the $Z_c(3885)$ 
and the $Z_c(3900)$ are in fact the same, the partial width for $D\bar{D}^*$ decays is $6.2\pm 2.9$ times larger than
that for $\jp\pi$.  This is small compared to open-charm versus hidden-charm decay-width ratios for established 
charmonium states that are above the open-charm threshold, such as the $\psi(3770)$ and $\psi(4040)$, where corresponding
ratios are measured to be more than an order-of-magnitude larger~\cite{PDG}. 

A study of $\pipi h_c(1P)$ final states in data taken at $\sqrt{s}=4.26$~GeV and nearby energy points
uncovered the sharp peak in the $M_{\rm max}(h_c\pi^{\pm})$ distribution near $4020$~MeV that is shown
in the center-right panel of Fig.~\ref{fig:zc}.   This peak, called the $Z_c(4020)$, has a fitted mass of
$M_{Z_c(4020)}=4022.9\pm 1.8$~MeV, which is only $\sim 6$~MeV above $2m_{D^*}$, and a width of
$\Gamma_{Z_c(4020)}= 7.9\pm 3.8$ ~MeV~\cite{bes_z4020}. This state looks like a charmed-sector version of the
$Z_b(10650)$.

A study of $\ee\rt D^*\bar{D}^*\pi^{\pm}$ events in the $\sqrt{s}=4.26$~GeV data sample produced the $M(D^*\bar{D}^*)$
invariant mass distribution shown in the right panel of Fig.~\ref{fig:zc}.  Here the limited available 
phase space limits the ability to see a distinct peak above background.  However, the distribution cannot be
explained by a phase-space term (dash-dot curve) plus background (dotted curve).  The best fit includes a
a BW term with peak mass near 4025~MeV, dubbed the $Z_c(4025)$~\cite{bes_z4025}.  Here a fit using a
mass-independent-width BW line shape function modified by a phase space factor that accounts for the nearby $2m_{D^*}$
mass threshold yields a mass  $M_{Z_c(4025)}=4026.3\pm 4.5$~MeV, which is $\sim 10$~MeV above the $2m_{D^*}$ threshold,
and a width  $\Gamma_{Z_c(4025)}= 24.5 \pm 9.5$~MeV.   The mass of the $Z_c(4025)$ is consistent within errors with the
mass of the $Z_c(4020)$, but its width is about $2\sigma$ higher.  Width measurements could be effected by interference
with a coherent background so, although it seems likely that the $Z_c(4020)$ and $Z_c(4025)$ are different decay channels
of the same state, more studies are needed before a firm conclusion can be drawn.

\subsection{Comments}

The recent BESIII findings, taken together with previous experimental results, establishes a concentration of
charmoniumlike states crowding the $D\bar{D}^*$ and $D^*\bar{D}^*$ mass threshold regions.  There are at least
four states very close the $D\bar{D}^*$ threshold.  The $X(3872)$ is right at the ${D^0} + {D^{*0}}$ threshold
and seems to be an isospin singlet~\cite{choi_prd}.  The $Z_c(3885)$ pole mass is about 10~MeV above the $D^0D^{*-}$ or
$D^+\bar{D}^{*0}$ threshold and is an isospin triplet.  Both have $J^P=1^+$ and couple to $S$-wave $D\bar{D}^*$ final
states.  The $Z_c(4020)/Z_c(4025)$ states show the existence of at least one isospin triplet just above the $D^*\bar{D}^*$
threshold.   If this is the charmed-sector equivalent of the $Z_b(10650)$, then it has $J^P=1^+$, in which case 
there is, at present, no obvious candidate for a isospin-singlet counterpart.  If such a state exists with
a mass that is above the $2m_{D^*}$ threshold and with a relatively narrow width, it might be accessible
in $B^-\rt K^- D^*\bar{D}^*$ decays.  BaBar has reported large branching fractions for 
$B^-\rt K^-D^{0*}\bar{D}^{*0}$ ($1.1\pm 0.1$\%) and $K^- D^{*+}D^{*-}$  ($0.13\pm 0.02$\%) but did not publish
any invariant mass distributions~\cite{babar_kdstrdstr}.  

The CMS group searched for a $b$-quark-sector equivalent
of the $X(38272)$ in the inclusive $\pipi\Upsilon(1S)$ invariant mass distribution produced in in proton-proton
collisions at $\sqrt{s}=8$~TeV, but found no evidence for peaks other than those due to $\Upsilon(2S)$ and
$\Upsilon(3S)$ to $\pipi\Upsilon(1S)$ transitions~\cite{cms_xb}.  However, if, as expected, the $b$-quark-sector
equivalent of the $X(3872)$ has $J^{PC}=1^{++}$, zero isospin, and is near the $B\bar{B}^*$ mass threshold, the
$\pipi\Upsilon(1S)$ decay mode, for which the $\pipi$ would originate from $\rho\rt\pipi$, would violate isospin and
be strongly suppressed relative to decays to the isospin-conserving $\omega\Upsilon(1S)$ final state. (This is not
the case for the $X(3872)$ where the isospin-allowed  $\omega\jp$ decay mode is kinematically suppressed: i.e.,
$Q_c\simeq m_{D^0}+m_{D^{*0}}-m_{\jp}=776$~MeV, which is twice the $\omega$ natural width below its peak mass
$m_{\omega}=783$~MeV.  In the $b$-quark-sector, $m_{B}+m_{B^*}-m_{\Upsilon(1S)}=1145$~MeV, which is well above $m_{\omega}$.) Thus,
$\omega\Upsilon(1S)$ final states are probably more relevant that $\pipi\Upsilon(1S)$ for searches for isospin-singlet
counterparts of the $Z_b(10610)$ and $Z_b(10650)$. This would require studies of decay final states that contain a
$\pi^0$, which will may be difficult to do with existing experiments, but could be done at BelleII~\cite{belleii}.

\section{Summary}

The QCD exotic states that are much preferred by theorists, such as pentaquarks, the $H$-dibaryon, and
meson hybrids with exotic $J^{PC}$ values continue to elude confirmation even in experiments with increasingly 
high levels of sensitivity.  On the other hand, a candidate $p\bar{p}$ bound state and a rich spectroscopy of
quarkoniumlike states that do not fit into the remaining unassigned levels for $\ccbar$ charmonium and $\bbbar$
bottomonium states has emerged.  No compelling theoretical picture has yet been found that provides a compelling
description of what is seen, but, since at least some of these states are near $D^{(*)}\bar{D}^*$ or $B^{(*)}\bar{B}^*$
thresholds and couple to $S$-wave combinations of these states, molecule-like configurations have to be important
components of their wavefunctions~\cite{braaten}.  This has inspired a new field of ``flavor chemistry'' that is
attracting considerable attention both by the experimental and theoretical hadron physics communities~\cite{nora}.

\section{Acknowledgements}

I thank the organizers of LEAP2013 for their hospitality and congratulate them on hosting such an interesting
and informative meeting.  While preparing this report I benefited from communications from E.~Braaten, P.~Ko
and Q.~Zhao and my colleagues on the Belle and BESIII experiments.  This work was supported by the Korean National
Research Foundation Grant No. 2011-0029457.


\begin{thebibliography}{99}
\bibitem{errors}  The errors I quote in this report are the statistical
and systematic errors added in quadrature.  For the actual numbers
please refer to the cited papers.
\bibitem{conj} In the rest of this report, the inclusion of
charge-conjugate modes is always implied.
\end{thebibliography}
\end{document}